%% file: paper-jlamp.tex
\documentclass{elsarticle}
\usepackage{ifpdf}
\usepackage{lineno} 
\usepackage{wrapfig}
\usepackage{cutwin}
\usepackage{multirow}
\usepackage{booktabs}
\usepackage[tight,footnotesize]{subfigure}
\usepackage{tweaklist}
\renewcommand{\enumhook}{
  \setlength{\topsep}{0em}
  \setlength{\itemsep}{0em}
  \setlength{\partopsep}{0em}
  \setlength{\parskip}{0em}
  \setlength{\parsep}{0em}
  \addtolength{\leftmargin}{-0.75em}
}
\renewcommand{\itemhook}{
  \setlength{\topsep}{0em}
  \setlength{\itemsep}{0em}
  \setlength{\partopsep}{0em}
  \setlength{\parskip}{0em}
  \setlength{\parsep}{0em}
  \addtolength{\leftmargin}{-0.75em}
}
\usepackage{placeins}

\ifpdf
\usepackage[colorlinks=true,linkcolor=blue,citecolor=blue]{hyperref}
\fi

\usepackage{graphicx}
\usepackage[cmex10]{amsmath}

\usepackage{array}

\usepackage{mdwmath}
\usepackage{mdwtab}

\usepackage{eqparbox}

\usepackage[tight,footnotesize]{subfigure}

\usepackage{url}
\usepackage{color}
\usepackage{xspace}

\usepackage{amsthm}

\makeatletter
\def\els@aparagraph[#1]#2{\elsparagraph[#1]{#2\@addpunct{.}}}
\def\els@bparagraph#1{\elsparagraph*{#1\@addpunct{.}}}
\makeatother

\newcommand\ADDoS{\textrm{ADDoS}\xspace}
\newcommand\seven{\textrm{SeVen}\xspace}

\newcommand\Factor{\mathtt{PMod}\xspace}
\newcommand\Pmod{\mathtt{PMod}}

\newcommand{\eg}{{\em e.g.}}
\newcommand{\ie}{{\em i.e.}}
\newcommand{\etal}{\emph{et al.}}

\newcommand{\Bscr}{\mathcal{B}}

\newcommand{\tup}[1]{\langle#1\rangle}

\newcommand\wait{\mathsf{WAITING}}
\newcommand\Wait{\mathsf{WAIT}}
\newcommand\In{\mathsf{IN}}
\newcommand\incall{\mathsf{INCALL}}

\newcommand\undef{\mathsf{undef}}

\renewcommand\cite[1]{\citep{#1}}

\newcommand{\coordinatedcall}{Coordinated Call\xspace} 

\journal{Journal of Logical and Algebraic Methods in Programming}

\begin{document}

\begin{frontmatter}

\author{Marcilio O. O. Lemos}
\ead{marciliolemos@ci.ufpb.br}
\address{Federal University of Para\'iba, Jo\~ao Pessoa, Brazil.}

\author{Yuri Gil Dantas}
\ead{dantas@mais.informatik.tu-darmstadt.de}
\address{Technische Universit\"at Darmstadt, Darmstadt, Germany}

\author{Iguatemi E. Fonseca}
\ead{iguatemi@ci.ufpb.br}
\address{Federal University of Para\'iba, Jo\~ao Pessoa, Brazil.}

\author{Vivek Nigam}
\ead{vivek.nigam@gmail.com}
\address{Federal University of Para\'iba, Jo\~ao Pessoa, Brazil.}


\title{On the Accuracy of Formal Verification of Selective Defenses for TDoS Attacks}

\begin{abstract}
Telephony Denial of Service (TDoS) attacks target telephony services, such as Voice over IP (VoIP), not allowing legitimate users to make calls. There are few defenses that attempt to mitigate TDoS attacks, most of them using IP filtering, with limited applicability. In our previous work, we proposed to use selective strategies for mitigating HTTP Application-Layer DDoS Attacks demonstrating their effectiveness in mitigating different types of attacks.
Developing such types of defenses is challenging as there are many design options, \eg, which dropping functions and selection algorithms to use. Our first contribution is to demonstrate both experimentally and by using formal verification that selective strategies are suitable for mitigating TDoS attacks.
We used our formal model to help decide which selective strategies to use with much less effort than carrying out experiments. Our second contribution is a detailed comparison of the results obtained from our formal models and the results obtained by carrying out experiments. We demonstrate that formal methods is a powerful tool for specifying defenses for mitigating Distributed Denial of Service attacks allowing to increase our confidence on the proposed defense before actual implementation.
\end{abstract}

\end{frontmatter}


\section{Introduction}

\input{structure/introduction}

\section{VoIP Protocols and the \coordinatedcall Attack}

\input{structure/voip-and-ddos-attacks}

\section{Selective Strategies}
\label{sec:seven}

\input{structure/seven}

\section{Simulations}
\label{sec:sim}

\input{structure/simulations}

\section{Experiments}
\label{sec:exp}

\input{structure/exp}

\section{Comparison between Simulation and Experimental Results}
\label{sec:comp}

\input{structure/comparison}

\section{Related and Future Work}
\label{sec:conc} 

\input{structure/future-work}

\paragraph{Acknowledgments}
This work has been funded by the DFG as part of the project Secure Refinement of Cryptographic Algorithms (E3) within the CRC 1119 CROSSING, by RNP project GT-ACTIONS, by Capes Science without Borders grant 88881.030357/2013-01 and CNPq.

\bibliographystyle{elsarticle-num}
\bibliography{master}

\end{document}

%% file: structure/introduction.tex
Telephony Denial of Service (TDoS) attacks is a type of Denial of Service (DoS) attack that target telephony services, such as Voice over IP (VoIP). With the increase in the popularity of VoIP services, we have witnessed an increase in TDoS attacks being used to target hospital line systems~\cite{Boston,NY} and systems for emergency lines (like the American 911 system)~\cite{KF}. According to the FBI, 200 TDoS attacks have been identified only in 2013~\cite{NY}. 

This paper investigates the use of \emph{selective defenses}~\cite{dantas14eisic} for mitigating one type of TDoS attack called \coordinatedcall~\cite{co-call} attack. The \coordinatedcall\ attack~\cite{co-call} exploits the fact that pairs of  attackers, Alice and Bob, can collude to exhaust the resources of the VoIP server. Assume that Alice and Bob are valid registered users.\footnote{This can be easily done for many VoIP services.} The attack goes by Alice simply calling Bob and trying to stay in the call as long as she can. {Since the server allocates resources for each call, by using enough pairs of attackers, attackers can exhaust the resources of the server denying service to honest participants. This is a simple, but ingenious attack, as only a relatively low rate of incoming calls is needed generating a small network traffic (when compared to SIP flooding attack for example)}. Thus it is hard for the network administrator to detect and counter-measure such attack. 

Formal methods and, in particular, rewriting logic can help developers to design defenses for mitigating DDoS attacks. In our previous work~\cite{dantas14eisic} we used selective strategies in the form of the tool \seven for mitigating HTTP Low-Rate Application-Layer DDoS attacks targeting web-servers. We formalized different attack scenarios in Maude~\cite{clavel-etal-07maudebook} and since our strategies are constructed over some probability functions, we used statistical model checking~\cite{sen05cav}, namely PVeStA~\cite{alTurki11calco}, to validate our defense. Due to our reasonable preliminary results, we implemented \seven\ and carried out experiments over the network obtaining similar results to the ones obtained using formal methods. It took us \emph{only 3 person months} to obtain our results using formal methods, while it took us \emph{24 person months} to obtain our first experimental results. Specifying scenarios using formal verification amounts to coding some few hundred lines of specification, while carrying out such experiments on the network amounts to building complex prototypes to carry out attacks, generate legitimate traffic and deploy defenses, integrating them with existing machinery, such as VoIP servers, setting and configuring the network and testing which take much more effort involving a larger team. Although we strongly believe that systems should also be validated by means of experiments, the confidence acquired from our formal analysis was invaluable for the success of this project.\footnote{Notice that although our experiments on the network were controlled experiments, they used off-the-shelf tools, such as Apache web-servers, which implement a number of optimizations not modeled in our formal specification. Moreover our experiments suffered from interference that cannot be controlled, such as network latency. The same is true for our results involving the VoIP server Asterisk used in our experimental results.}

This paper provides more evidence supporting the claim that formal methods can help specifiers in designing selective defenses. We systematically consider a number of selective defenses used for mitigating TDoS attacks. We  compare the results obtained using our formal specification and the results obtained implementing such defenses and carrying out experiments on the network. Our results show a high accuracy for most of the results, specially on availability, but less accurate on results involving time measurements. 

Our contributions are three-fold:
\begin{itemize}
  \item We formalized in Maude the \coordinatedcall attack and three selective defenses based on \seven: the first using a uniform selection strategy, the second with roulette selection strategy~\cite{roulette}, and the third with a tournament selection strategy~\cite{Blickle:1995:MAT:645514.658088}. We also considered two models for legitimate call duration: an exponential call duration which models traditional telephony~\cite{Brown_2002a} and lognormal call duration which models VoIP telephony~\cite{7414101}.

   We carried out a number of simulations using PVeStA to test the efficiency of each version the defense used under the two different assumptions on call duration. Our simulation results suggest that \seven mitigates the \coordinatedcall attack;

  \item We implemented the different selective defenses analysed using our formal models, and integrated them with the VoIP server Asterisk~\cite{asterisk} using the SIP-protocol. We also implemented the \coordinatedcall attack. Our experimental results demonstrate in practice that our selective defenses can mitigate the \coordinatedcall attack;

  \item Finally, we compare the results obtained from our formal analysis with the results obtained from our experimental results to analyze the accuracy of the results obtained from our formal analysis. This comparison demonstrates that formal methods are of great value as they can be used early on to develop and evaluate new defense mechanisms for mitigating TDoS attacks with much less effort than implementing defenses and carrying out experiments on the network.
\end{itemize}

A small subset of experimental and simulation results appearing in this paper appeared in our previous work~\cite{lemos16sbrc,dantas16wrla} which only considered scenarios where call duration followed a uniform probability and a single mechanism for dropping calls, namely the roulette strategy. This paper extends our previous work by considering different assumptions on call duration, namely lognormal distribution, modeling usual VoIP calls, and exponential distribution, modeling usual telephony, \ie, non VoIP calls. Moreover, we consider here different mechanisms for dropping calls, namely uniform, roulette and tournament dropping strategies. In terms of total time of experiments, the results in this paper add more than 40 hours of experimental results when compared to the results in our previous work~\cite{lemos16sbrc,dantas16wrla}.

This paper is organized as follows. Section~\ref{sec:voip} we review the Session Initiation Protocol (SIP) used for initiating a VoIP call and also explain the \coordinatedcall\ attack. Section~\ref{sec:seven} describes how \seven\ works, while Section~\ref{sec:formal} details its formalization in Maude. Sections~\ref{sec:sim} and \ref{sec:exp} contain our simulation and experimental results including our main assumptions, results and discussion of the results obtained and Section~\ref{sec:comp} discusses the accuracy of our simulation results. We discuss in Section~\ref{sec:conc} related and future work. The implementation used to carry out our simulations is available for download at \cite{seven}.

%% file: structure/voip-and-ddos-attacks.tex
\label{sec:voip}
\begin{figure}[t]
  \vspace{-4mm}
	\begin{center}
	\includegraphics[width=8cm]{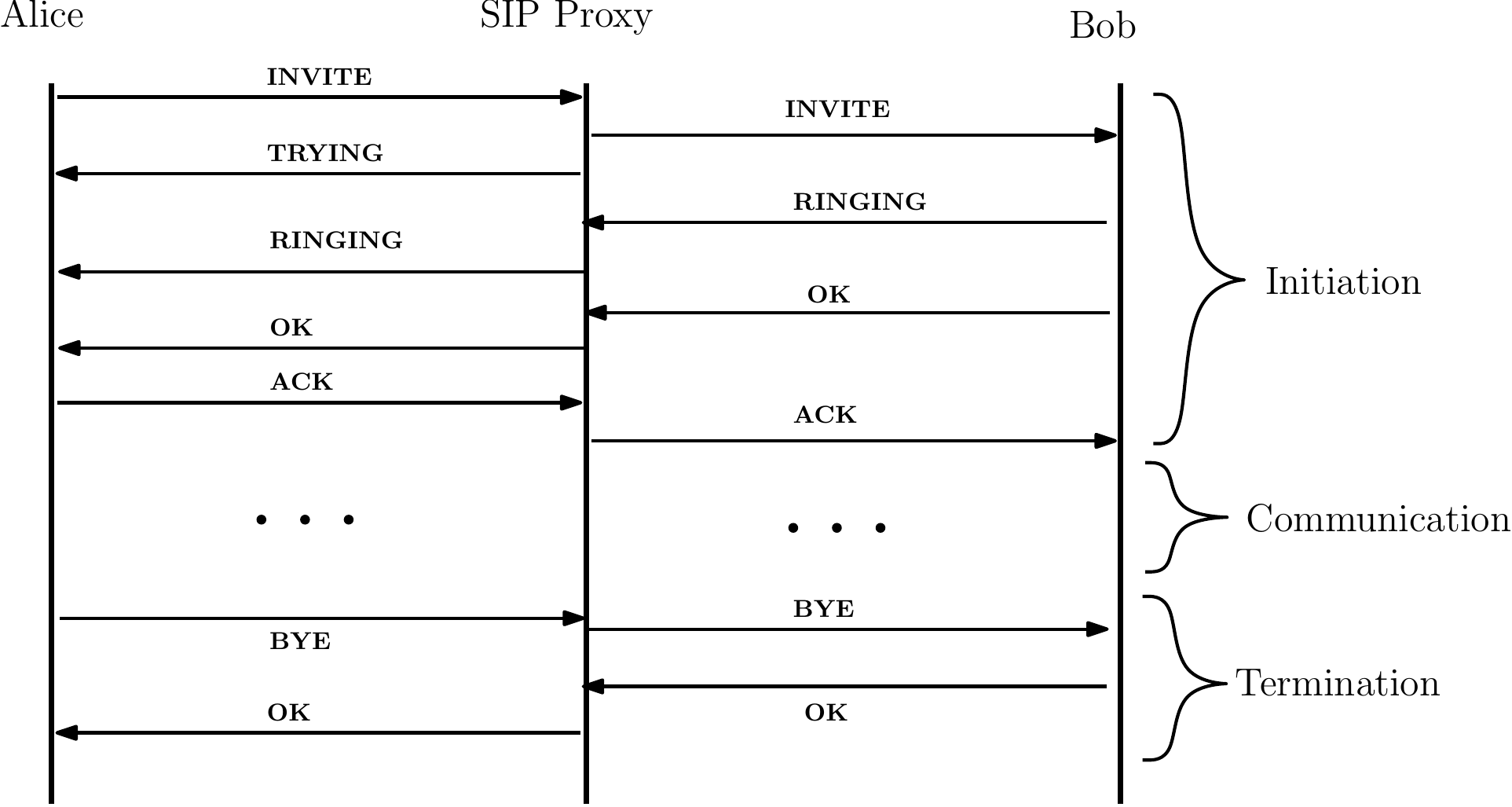}
	\end{center}
  \vspace{-4mm}
	\caption{Exchange of messages between the server and two users (Alice and Bob) during a normal execution of the SIP protocol.}
	\label{fig:sip-normal}
	\vspace{-4mm}
\end{figure}
We now review the Session Initiation Protocol~\cite{sip}, which is one of the main protocols used to establish Voice over IP (VoIP) connections. 
Figure~\ref{fig:sip-normal} shows the message exchanges performed to establish a connection between two registered users, Alice and Bob, where Alice tries to initiate a conversation with Bob. It also contains the messages exchanged to terminate the connection. 

For initiating a call, Alice sends an INVITE message to the SIP server informing that she wants to call Bob. If Bob or Alice is not registered as valid users, the server sends a reject message to Alice. Otherwise, the server sends an INVITE message to Bob.\footnote{In fact, we omit some steps carried out by the server to find Bob in the network. This step can lead to DDoS amplification attacks~\cite{shankesi09esorics} for which known solutions exists. Such amplification attacks are not, however, the main topic of this paper.} At the same time, the server sends a TRYING message to Alice informing her the server is waiting for Bob's response to Alice's invitation. 
The server waits for a RINGING message from Bob indicating that Bob’s telephone is ringing. Bob might reject the request, in which case the server informs Alice (not shown in the Figure), or accept the call by sending the message OK.
Finally, the server sends the message OK to Alice who sends an ACK message back to the server which forwards it to Bob.

At this point, the communication is established and Alice and Bob should be able to communicate as long as they need/want. (This is represented by the three ellipses in Figure~\ref{fig:sip-normal}.) The call is then terminated once one of the parties (Alice) sends a BYE message to the server.  The server then sends a BYE message to the other party (Bob), which then answers with the message OK, which is forwarded to Alice, and the connection is terminated.

\paragraph{Coordinated VoIP Attack~\cite{co-call}} 
\label{sub:new_attack}
A pair of colluding attackers, $A_1$ and $A_2$, that are registered in the VoIP service,\footnote{Or alternatively two honest users that have been infected to be zombies by some attacker.} call each other and stay in the call for as much time as they can. Once the call is established, the attackers stay in the call for indefinite time. They might be disconnected by some Timeout mechanism establishing some time bounds on the amount of time that two users might call. 
During the time that $A_1$ and $A_2$ are communicating, they are using resources of the server. Many VoIP servers have an upper-bound on the number of simultaneous calls they can handle. If enough pairs of attackers collude, then the resources of the server can be quickly exhausted. This attack is hard to detect using network analyzers because the traffic generated by attackers is similar to the traffic generated by legitimate clients. The attackers follow correctly the SIP protocol and, moreover, there is no need to generate a large burst of calls, \emph{but rather place calls in a moderate pace. Eventually, the server's capacity will be exhausted.}

There are many reasons why VoIP devices participate in a \coordinatedcall attack. Pairs of legitimate users may be unsatisfied with the VoIP provider and participate in such attacks. Attackers may also use botnets with some infected malware. There has been evidence of the use of botnets in 2007~\cite{VoIPBotnet}. Tools that can place and receive calls (SIPp~\cite{chico:bioca}) and tutorials on the Internet help automate the steps for carrying out the \coordinatedcall attack. 

Indeed, we have done so and as we demonstrate in Section~\ref{sec:exp}, the \coordinatedcall attack can reduce considerably availability to levels around 5\% without generating large amount of traffic.


%% file: structure/seven.tex
We proposed in our previous work~\cite{dantas14eisic} a new defense mechanism, called \seven, for mitigating Application-Layer DDoS attacks (\ADDoS) using selective strategies~\cite{6056588}. An application using selective strategies does not immediately process incoming messages, but waits for a period of time, $T_S$, called a round. {During a round, \seven\ accumulates messages received in an internal buffer. Normally, this internal buffer reflects the connections of the protected service. Assume that $k$ is the maximum capacity of the service being protected. For VoIP servers $k$ is the number of calls that the application can handle simultaneously. If the number of messages accumulated reaches $k$ (the size of internal buffer) and a new incoming request \emph{R} arrives, \seven behaves as follows:}

\begin{enumerate}
    \item \seven\ decides whether to process \emph{R} or not based on a probability \ensuremath{P_{1}}. $P_1$ is defined using the counter $\Factor$ following~\cite{khanna08infocom}:
    \[
    \frac{k}{k+\Factor}
    \]
    At the beginning of the round, we set $\Factor = 0$. $\Factor$ is incremented whenever the application's capacity is exhausted and a new incoming request arrives reducing thus the probability of new incoming request being selected by \seven\ during a round. The intuition is that $P_1$ reduces with the increase of incoming traffic thus reducing the impact of high numbers of request to the application;

    \item  If \seven\ decides to process \emph{R}, then as the application is overloaded, it should decide which request currently being processed should be dropped. This decision is governed by \ensuremath{P_2}, a distribution probability \emph{which might depend on the state of the existing request};  \label{step:2}
    \item Otherwise, \seven\ simply drops the request \emph{R} without affecting the requests currently being processed and sends a message to the requesting user informing that the service is temporarily unavailable.
\end{enumerate}

At the end of the round, \seven\ processes the requests that are in its internal buffer (surviving the selective strategy) and sends them to the application.

\seven\ mitigates attacks \emph{only when the maximum capacity of the VoIP server is reached}. When this happens, \seven\ has two mechanisms for dropping requests. The first one is by using probability $P_1$ and the other by using $P_2$. The main goal of the former is to mitigate the impact of volumetric attacks~\cite{6056588}. This is because whenever the defense receives a high volume attack the value of $\Factor$ increases rapidly increasing rapidly the chance of dropping a request. This is also reflected on the round time $T_S$ which is typically in the order of some hundred milliseconds to avoid $\Factor$ from reaching too high values even under normal traffic. We used $T_S = 100$ms.

As Coordinated Calls do not generate a very large number of requests, the mechanism using $P_1$ is not the main mitigation mechanism used by \seven\ for this attack, but rather the mechanism using $P_2$. 
There is, however, much space for specifying the probability distribution $P_2$ governing \seven. In \cite{dantas14eisic}, we showed that by using simple \emph{uniform distributions} for dropping existing requests, \seven\ can be used to mitigate a number of \ADDoS\ attacks using the HTTP protocol, such as the Slowloris and HTTP POST attacks even in the presence of a large number of attackers.


For mitigating the \coordinatedcall\ attack described in Section~\ref{sec:voip}, we set the probability \ensuremath{P_{2}}, governing which call to be dropped from the internal buffer, to depend on (1) the status of the call and (2) on the duration of a call. We consider two types of call status: 
\begin{itemize}
	\item $\wait$: A call is $\wait$ if it has already sent an INVITE message, but it is still waiting for the responder to join the call, that is, it has not completed the initiation part of the SIP protocol;
	\item $\incall$: A call is $\incall$ if the messages of initiation part of SIP have been completed and the initiator and the responder are already communicating (or simply in a call).
\end{itemize}
Thus, any incoming INVITE requests assume the status of $\wait$, and these can change they status to $\incall$ once the initiation part of SIP is completed.

We assume here that it is preferable to a VoIP server, when overloaded, to drop 
$\wait$ requests than $\incall$ requests that are communicating not for a \emph{very long duration}. In many cases, it is true that interrupting an existing call is considered to be more damaging to a server reputation than not allowing a user to start a new call. This could also be modeled by configuring the probability distributions of \seven\ accordingly.
To determine whether a call is taking too long, we assume that the server knows what is the average duration, $t_M$, of calls.\footnote{The value of $t_M$ can be obtained by the history of a VoIP provider's usage.}

\begin{figure}
	\vspace{-4mm}
	\begin{center}
	\includegraphics[width=6cm]{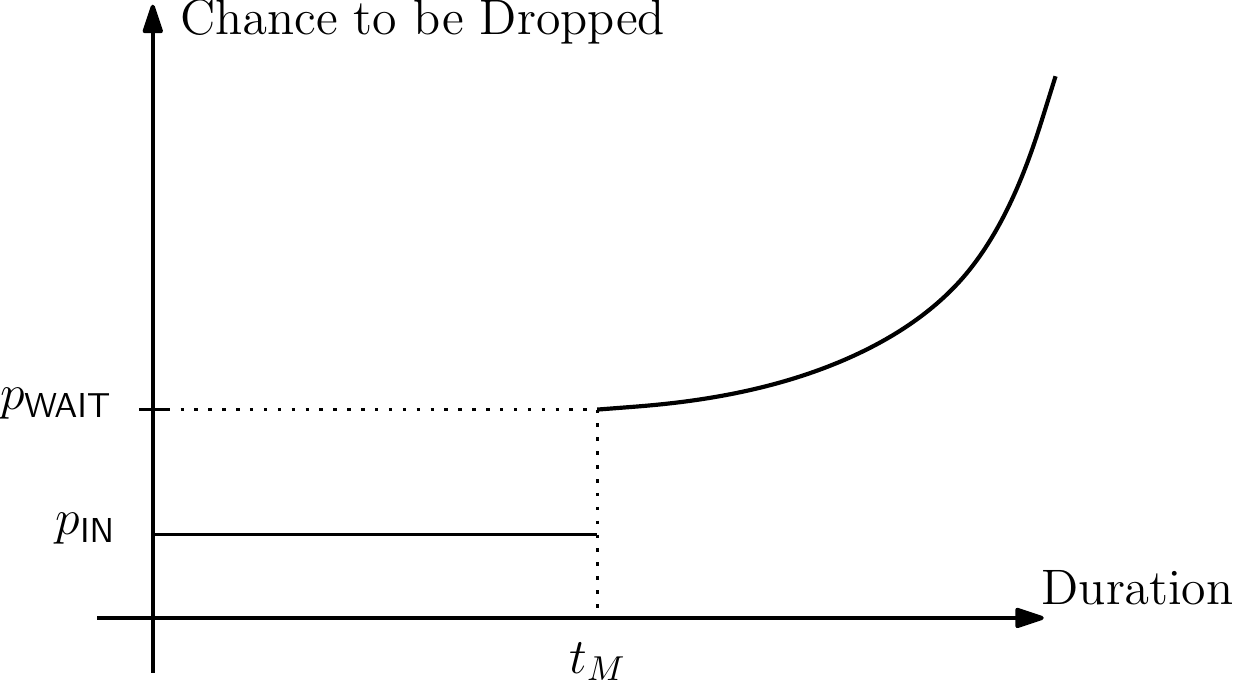}		
	\end{center}
  \vspace{-3mm}
	\caption{Graph (not in scale) illustrating the behavior of \seven\ according to the status of a call and its duration. $p_{WAIT}$ is the probability of dropping a $\wait$ call, while $p_{IN}$ the probability of dropping a $\incall$ call.}
	\label{fig:exp}
	\vspace{-4mm}
\end{figure}

The dropping factor of an $\incall$ request increases exponentially once the call duration is greater than $t_M$. Figure~\ref{fig:exp} depicts roughly the dropping factor used to drop requests. The actual function $d$ (for drop factor) is of the form, where $t$ is the call duration and $\alpha$ is a parameter:

\begin{equation}
d(t) = \left\{
\begin{array}{ll}
  p_{\textrm{WAIT}} & \textrm{if } t = 0\\
  p_{\textrm{IN}} & \textrm{if } 0 < t \leq t_M\\
  p_{\textrm{WAIT}} + e^{\alpha t /t_M} & \textrm{if } t > t_M
\end{array}\right.
\label{eq:drop}  
\end{equation}

Given this dropping factor, we consider in our analysis three ways for selecting which
call to drop. Assume the server has a capacity of $k$ simultaneous calls. Moreover, assume $c_1, \ldots, c_k$ are the calls currently being processed by the server and they have dropping factors of, respectively, $d_1,\ldots,d_k$. We considered three different selection strategies described in the literature, namely, uniform~\cite{dantas14eisic}, roulette~\cite{roulette} and $n$-tournament~\cite{Blickle:1995:MAT:645514.658088}:

\begin{itemize}
    \item \textbf{Uniform:} In this strategy, the dropping factor of a call is not considered. We select using uniform probability which call is going to be dropped. Thus any call independent on its duration and status can be selected to be dropped by \seven;

    \item \textbf{Roulette:}  In the roulette strategy~\cite{roulette}, we select randomly a call $c_i$ to be dropped where the probability of being dropped is proportional to its dropping factor. Thus in the roulette strategy a call $c_i$ has twice the chance of being dropped than a call $c_j$ if $d_i = 2 \times d_j$.

    For instance, consider $k=4$ and that the server is serving the calls $c_1, c_2, c_3, c_4$ with dropping factors $2,3,1,6$ respectively. We select using uniform distribution a number $r$ between $0$ and $2 + 3 + 1 + 6 = 12$. If $0 \leq r < 2$, then the call $c_1$ is selected, if $2 \leq r < 5$, then the call $c_2$ is selected, if $5 \leq r < 6$, then the call $c_3$ is selected, and otherwise if $6 \leq r < 12$ then the call $c_4$ is selected. In this way, the call $c_4$ has 6 times more chance to be selected than the call $c_3$ for example;

    \item \textbf{$n$-Tournament:} In the $n$-tournament strategy~\cite{Blickle:1995:MAT:645514.658088}, we first select $n$ calls randomly using uniform probability to be part of the \emph{tournament}. Then, the call to be dropped will be the call with the greatest dropping factor among the $n$ selected calls. In case there is more than one possible call with the greatest dropping factor, we select one of them at random.

    For instance, in the example above, if $n=2$, then we would select randomly two out of the four calls $c_1, c_2, c_3, c_4$ to participate in the tournament. Say the calls $c_2$ and $c_3$ are chosen to be part of the tournament. In this case, the call $c_2$ is selected to be dropped as it has the greatest drop factor.

    Notice that if $n$ is chosen to be too low when compared to $k$, the $n$-tournament behaves closer to the uniform dropping strategy. In fact, if $n=1$, then the $n$-tournament can be shown to be equivalent to the uniform dropping strategy.
    On the other hand, if $n$ is chosen to be too high, then the $n$-tournament behaves closer to a deterministic dropping strategy that selects the call with the greatest dropping factor. Indeed, if $k = n$, then the $n$-tournament strategy is deterministic. 
    In our experiments, we used $n = k / 2$, that is, a strategy between the uniform and a deterministic strategy.
\end{itemize}

While the attackers attempt to stay in a call for very long periods of time, legitimate clients do not behave so. The literature models legitimate call duration using the following distributions, where the parameters $\lambda, \sigma$ and $\mu$ are computed accordingly to the mean call durations assumed $t_M$ (see~\cite{jewett80book} for more details):

\begin{itemize}
     \item \textbf{Exponential:} Typical telephony models~\cite{Brown_2002a}, \ie, not VoIP, assume that the call duration of legitimate clients follows an exponential density distribution:
     \begin{equation}
     \label{eq:exp}
     f(x,\lambda) = \left\{
     \begin{array}{ll}
       \lambda e^{-\lambda x} & x \geq 0\\
       0 & x < 0
     \end{array}\right.      
     \end{equation}
     Since the coordinated call attack can also be carried out in standard telephony systems, we considered call duration following this distribution.

     \item \textbf{Lognormal:} While standard telephony calls are paid per duration, in VoIP calls have fixed rates or are even for free. This difference impacts legitimate call duration which in VoIP follows a lognormal density distribution~\cite{7414101}:
     \begin{equation}
      \label{eq:lognormal}
       f(x,\mu,\sigma) = \frac{1}{x\sigma\sqrt{2\pi}} 
                       \mathrm{exp}\left[- \frac{(\ln(x) - \mu)^2}{2\sigma^2} \,\right]\,.
     \end{equation}
\end{itemize}

\subsection{Sample Execution}

Consider the following application state, $\Bscr_i$, at the beginning of a round and assume that $k = 3$, $\Factor=0$, the current time is $9$ and the average call duration is $t_M = 5$ time units:
\[
  \Bscr_1 = [\tup{id_1, \wait, \undef}, \tup{id_2, \incall, 0.5}]\\ 
\]
$\tup{id,st,tm}$ specifies that the call $id$ has status $st$ and the call started at time $tm$ where $tm$ is $\undef$ whenever $st = \wait$.
This buffer specifies that $id_1$ is waiting the responding party to answer (with a OK message) his invitation request and that $id_2$ is currently in a call.  This means that $id_2$ is calling already for way more than the expected average. 

Assume that a message $\tup{id_1, \textrm{OK}}$ at time $9.5$ arrives 
specifying that the responder of the request $id_1$ answered the call. The buffer is updated to the following:
\[
  \Bscr_2 = [\tup{id_1, \incall, 9.5}, \tup{id_2, \incall, 0.5}]\\ 
\]

Then the message $\tup{id_3, \textrm{INVITE}}$ arrives. Since $k$ has not yet been reached, a new request is inserted in the buffer and the message TRYING is sent to the requesting user. Notice that the RINGING message is not yet sent to the responding user. The buffer changes to:
\[
  \Bscr_3 = [\tup{id_1, \incall, 9.5}, \tup{id_2, \incall, 0.5}, \tup{id_3, \wait, \undef}]    
\]

Suppose now that another message $m_1 = \tup{id_4, \textrm{INVITE}}$ arrives at time $10.5$. As the buffer is now full, it sets $\Factor$ to $1$ and the application has to decide whether it will keep $m_1$. \seven generates a random number in the interval [0,1] using uniform distribution. Say that this number is less than $3/(3 +1)$, which means that it will select to process $m_1$. 
However, it has to drop some existing request. The current requests $id_1,id_2,id_3$ have dropping factors following Figure~\ref{fig:exp}:
\begin{itemize}
    \item $id_1$ has dropping factor $p_{\mathsf{IN}}$ to be dropped  because it is calling for a duration less than $t_M$: $10.5 - 9.5 < 5$;
    \item $id_2$ has a much higher dropping factor because it is calling for twice $t_M$: $10.5 - 0.5 = 2 \times 5$;
    \item $id_3$ has dropping factor of $p_{\mathsf{WAIT}}$ as it has $\wait$ status.
\end{itemize}
The application decides which one to drop either using \emph{uniform probability}, in which case the dropping factor of requests is not considered, or the \emph{roulette strategy}, in which case $id_2$ has a greater probability of being dropped, or the $n$-tournament strategy in which case it would depend on $n$. 

Suppose that the application decides to drop $id_2$, which means that the call is interrupted by the application. The resulting buffer is:
\[
     \Bscr_4 = [\tup{id_1, \incall, 9.5}, \tup{id_4, \wait, \undef}, \tup{id_3, \wait, \undef}]
\] 
Assume that now the round time is elapsed. The application sends a RINGING message to the responder of the requests $id_3$ and $id_4$.

\section{Formal Specification} 
\label{sec:formal}

Our specification follows~\cite{dantas14eisic,eckhardt12fase,eckhardt12wadt} by specifying test scenarios using actors where attackers, clients, and the server send and receive messages. These messages are stored in a scheduler that maintains a queue of messages.
The attackers do not take control over the channel. Instead they share a channel with the clients. 

We formalize all actors in Maude~\cite{clavel-etal-07maudebook} and carry out simulations by using the statistical model checker PVeStA~\cite{alTurki11calco}.  For simplicity, we considered the server and \seven as one actor, which means that \seven is also able to operate as a normal SIP Server, \eg, processing and establishing call connections. Such decision does not affect the analysis of our results as in practice \seven and the VoIP server are in the same machine and thus share a quick communication channel. In the following, we describe our Maude specification. The complete model can be found in~\cite{seven}.

We refer to~\cite{clavel-etal-07maudebook,DBLP:journals/jlp/Meseguer12} for a more detailed description of Maude and its underlying foundations on Rewriting Logic.

\subsection{Key Sorts and Functions}

\paragraph{Actor}
The elements of the sort \texttt{Actor} is constructed by the operator
\begin{verbatim}
op <name:_|_> : Address AttributeSet -> Actor .
\end{verbatim}
which takes an \texttt{Address}, which can be a string, and a set of attributes, 
\texttt{AttributeSet}. In our formalization, we use the following attributes:
\begin{verbatim}
op req-cnt:_ : Float -> Attribute .
op b-set:_ : NBuffer -> Attribute .
op server:_ : Address -> Attribute .  
op status:_ : Status -> Attribute
\end{verbatim}
The attribute \texttt{req-cnt} stores the value of $\Pmod$, \texttt{b-set} stores the 
internal buffer of the server of sort \texttt{NBuffer} and \texttt{server} stores the 
address of the server. Finally, the attribute \texttt{status} specifies the status of the call which may be any of the following state constants: 
\begin{itemize}
  \item \texttt{none} -- a call that has not yet been placed;
  \item \texttt{invite} -- a call that has been placed and is waiting for the responder to answer;
  \item \texttt{incall} -- a call where the participants are currently communicating;
  \item \texttt{complete} -- a call that has been completed, \ie, the parties have communicated for the expected time;
  \item \texttt{incomplete} -- a call that was interrupted while communicating by \seven.
\end{itemize}

The \seven buffer has sort \texttt{NBuffer} and is constructed by pairing a number and a buffer:
\begin{verbatim}
op [_|_] : Nat Buffer -> NBuffer . 
\end{verbatim}
The number specifies the elements in the buffer which is a list of elements of sort \texttt{EleBuf}. These elements are constructed using a 3-tuple of the form:
\begin{verbatim}
op <___> : Address State Float -> EleBuf .
\end{verbatim}
The first position stores the address of the call. The second position denotes the state of the call. The third position stores the time of the first request of the call and is used to compute the call duration. 

\paragraph{Messages}
Actors process messages of sort \texttt{Msg} which are constructed using the following 
operator
\begin{verbatim}
op _<-_ : Address Contents -> Msg . 
\end{verbatim}
The first parameter of sort \texttt{Address} specifies to which actor the message is 
directed and the second parameter of sort \texttt{Contents} is the payload of the message.
An \texttt{ActiveMsg} is a timestamped \texttt{Msg} constructed using the following operator:
\begin{verbatim}
op {_,_} : Float Msg -> ActiveMsg .
\end{verbatim}
The first parameter specifies the time when the paired message is to be processed.

We assume that an active message \texttt{\{gt1,msg1\}} is always going to be processed before an active message 
\texttt{\{gt2,msg2\}} whenever \texttt{gt1 < gt2}. This is accomplished by using a Scheduler as in~\cite{dantas14eisic,eckhardt12fase,eckhardt12wadt}. A scheduler has sort \texttt{Scheduler} and is constructed by the following operator
\begin{verbatim}
op [_|_] : Float ActiveMsgList -> Scheduler .
\end{verbatim}
The first parameter is the global time while the second parameter contains the list of active messages ordered by their delivery time. The following function returns the scheduler obtained by inserting at the correct position, \ie, according to the messages timestamp, a list of active messages into a given scheduler. 
\begin{verbatim}
op insert : Scheduler ActiveMsgList -> Scheduler .  
\end{verbatim}

\paragraph{Configuration}
A configuration of sort \texttt{Config} is a collection of \texttt{Actors} and \texttt{Scheduler}:
\begin{verbatim}
subsort Actors < Config .
subsort Scheduler < Config .
op __ : Config Config -> Config [assoc comm id: none] .
\end{verbatim}

For example, the initial configuratiom is defined by the equation:
\begin{verbatim}
eq initState = 
	<name: server | req-cnt: 0.0 , b-set: [0 | none], none > 
	<name: client-generate | server: server, cnt: 0 , none >
	<name: attacker-generate | server: server, cnt: 0 , none >
        [0.0 | {0.0, (attacker-generate <- spawn )} ;
               {0.0, (client-generate <- spawn )} ;
               {Ts, server <- ROUND}] .
\end{verbatim}
It specifies a configuration with three actors: an attacker generator, a client generation and a server. The global time is \texttt{0.0} and there are three active messages in the scheduler, the first two directed to the actors \texttt{attacker-generate} and \texttt{client-generate}, respectively. Here we follow the Shared Channel Model~\cite{6056588} where clients and attackers share the same channel. Thus the application does not distinguish malicious traffic from legitimate one as it is usually the case, in particular, for the Coordinated Call attack.

Intuitively, the user specifies the rate at which new client and attacker actors are created. Then, for instance, when the message \texttt{client-generate} is processed a new client actor is created and a new message \texttt{client-generate} is scheduled so that a new client actor is created according to the rate given. Similarly, when processing \texttt{attacker-generate} which creates a new attacker actor and a new \texttt{attacker-generate} is scheduled to be processed according to the attacker generating rate. These rewrite rules are omitted.

The third message scheduled at the time \texttt{Ts} is directed to the \texttt{server} which is implementing the \seven strategy establishing when a \seven round ends.

The following function extracts the first scheduled active message in a scheduler and returns a 
new scheduler with the global time advanced to its delivery time. 
\begin{verbatim}
 op mytick : Scheduler -> Config .
\end{verbatim}
For instance, let \texttt{msg} be a message and \texttt{SL} an \texttt{ActiveMsgList}. Then
\begin{verbatim}
  mytick([0.0 | {1.0,msg} ; SL]) 
\end{verbatim}
returns \texttt{msg [1.0 | SL]} containing the message \texttt{msg} and the scheduler \texttt{[1.0 | SL]}. Intuitively, the message \texttt{msg} is going to be processed next. Message processing is formalized by rewriting rules.

\paragraph{Selective Strategies}
Finally, we specified the three selective strategies described in Section~\ref{sec:seven}, namely, Uniform, Roulette and Tournament. The function
\begin{verbatim}
op select : Float Buffer -> ActorBuffer .  
\end{verbatim}
implements one of the selective strategies. Given the global time and a buffer, this function returns a pair of sort \texttt{ActorBuffer} with the actor name of the selected element that has been selected to be removed and the resulting buffer obtained by removing this element from the given buffer.

All selective strategies we formalized use the following function:
\begin{verbatim}
op sampleUniWithInt : Nat -> Nat
\end{verbatim}
which for a given input $n$ returns a random natural number between $0$ and $n$. The following 
function uses this function to select at random an element from a given buffer and thus to implement the uniform selection strategy. 
\begin{verbatim} 
op selectRandom : Buffer -> EleBuf .
\end{verbatim}

For the roulette strategy, we compute using the dropping factor 
\begin{verbatim}
op roulette : Float Buffer -> EleBuf .
\end{verbatim}
which creates a roulette by assigning weights to the elements of the buffer according to the dropping factor and then randomly selects one.

A tournament for the $n$-tournament selection strategy is created by the following function:
\begin{verbatim}
op creatingTour : NBuffer Nat Buffer -> Buffer .
\end{verbatim}
It takes an NBuffer and a natural number, specifying the size of the tournament, and accumulates 
the selected elements to the tournament in the third argument. Once the tournament is created, we use the following function to select the one with the greatest dropping factor:
\begin{verbatim}
op selectGreatest : Buffer -> EleBuf .
\end{verbatim}
which takes a buffer with the tournament traversing it to find the element with the greatest dropping factor. 
\subsection{Rewrite Rules}

The rewrite rules modify elements from \texttt{Conf} and specify the operational semantics of a system. We describe next the main rewrite rules used in our formalization. 

The first action we describe is when an actor receives a \texttt{poll} message indicating that it should start a call at time \texttt{gt + delay}.
\begin{verbatim}
rl [CLIENT-RECEIVE-POLL] : 
  <name: c(i) | server: Ser, status: none, AS >
  {c(i) <- poll} [gt | SL]
 => 
  <name: c(i) | server: Ser , status: invite, AS >
  mytick(insert([gt | SL], { gt + delay, Ser <- INVITE(c(i))})) .
\end{verbatim}
The following rewrite rule specifies the behavior of a client upon receiving a RINGING message from the server. It changes the client's state from \emph{invite} to \emph{connected} and generates a message BYE, scheduled to be sent after some time. This means that all legitimate clients do not overpass the average time of the duration of calls using one of the call duration models, exponential or lognormal, described in Section~\ref{sec:seven}. This means that the client called \texttt{c(i)} is expected to end its call at time \texttt{gt + callDur(tMedio)}.  
\begin{verbatim}
rl [CLIENT-RECEIVE-RINGING] : 
  <name: c(i) | server: Ser, status: invite, AS >
  {c(i) <- RINGING} [gt | SL]
 => 
  <name: c(i) | server: Ser , status: connected, AS >
  mytick(insert([gt | SL], 
        { gt + callDur(tMedio), (Ser <- BYE(c(i)))})) .
\end{verbatim}
\seven may, however, drop a call before the call is finished. We classify such a call as an incomplete call, \ie\ the dropped client's status is changed to \texttt{incomplete}. We omit this rule.

The rewrite rules for the attackers are similar to the client rules. The only difference is that no \texttt{BYE} message is generated, thus, specifying the Coordinated Call attack where attackers attempt to stay in the call for indefinite time. We elide these rules.

\begin{figure}[t]
\begin{verbatim}
crl [SeVen-RECEIVE-INVITE] : 
    <name: Ser | req-cnt: pmod , b-set: [lenB | B],  AS > 
    {Ser <- INVITE(Actor)} [gt | SL]
 => if (lenB >= lenBufSeVen) then
          if p1 then ConfAcc myTick(SchAcc)
                else ConfRej myTick(SchRej)
          fi 
    else ConfAcc2 mytick(SchAcc2)
    fi 
if p1 := sampleBerWithP(accept-prob(pmod)) 
 /\ { ActorDr, bufDr } := select(gt,B)
 /\ nBuf := add([lenB + (- 1) | bufDr], < Actor invite gt >) 
 /\ ConfAcc := <name: Ser | req-cnt: (pmod + 1.0), b-set: nBuf , AS >
 /\ SchAcc := insert([gt | SL], 
                     {gt, Actor <- TRYING} ; {gt, ActorDr <- poll})
 /\ ConfRej := <name: Ser | req-cnt: (pmod + 1.0), b-set: [lenB | B], AS >
 /\ SchRej := insert([gt | SL], 
                     {gt + delay , Actor <- poll})
 /\ b-setNu := add( [lenB | B], < Actor invite gt > )
 /\ ConfAcc2 := <name: Ser | req-cnt: pmod , b-set: b-setNu, AS > 
 /\ SchAcc2 := insert([gt | SL], {gt + delay, Actor <- TRYING}) .

rl [SeVen-APP-ROUND] :
     <name: Ser | req-cnt: pmod , b-set: [lenB | B], AS >
     {Ser <- ROUND} [gt | SL]
 =>
     <name: Ser | req-cnt: 0.0, b-set: [lenB | B], AS >
     mytick(insert([gt | SL],  
               {gt, reply(Ser, B, gt)} {gt + Ts, Ser <- ROUND})) .
\end{verbatim}  
\caption{Rewrite rules specifying \seven's selective strategy.}
\label{fig:seven-rules}
\end{figure}

Figure~\ref{fig:seven-rules} depicts the rules implementing \seven's strategy. For each INVITE message received by some actor \texttt{Actor}, the rule \texttt{SeVen-RECEIVE-INVITE} checks whether the buffer of the server reached its maximum. If not, then the incoming request is added to the server's buffer (\texttt{ConfAcc2}) and a message TRYING to the corresponding actor is created. Otherwise, \seven throws a coin (\texttt{p1}) to decide whether the incoming request will be processed using \texttt{pmod}. If \seven decides to process the incoming request, then some request being processed is selected to be dropped using the function \texttt{select}. It returns the name of the actor \texttt{ActorDr} and the resulting buffer \texttt{nBuf}. The incoming request is added to \texttt{nBuf} and \texttt{pmod} gets incremented, resulting in the configuration \texttt{ConfAcc}. Moreover, a \texttt{poll} message to \texttt{ActorDr} and a \texttt{TRYING} to \texttt{Actor} are created specifying that the connection is going to be terminated. Otherwise, the incoming request is rejected and \texttt{pmod} is incremented without affecting the server's buffer resulting in the configuration \texttt{ConfRej}. A poll message to \texttt{Actor} is also created.

The rule \texttt{SeVen-APP-ROUND} specifies that when the round finishes, all surviving \texttt{WAITING} requests in \seven's buffer are answered by the function \texttt{reply}. Then a new round starts and \texttt{pmod} is re-set.

%% file: structure/simulations.tex
We detail our simulation results obtained from our formal specification using the statistical model checker PVeStA~\cite{alTurki11calco}. Our simulations are parametric in the following values:

\begin{itemize}
	\item \textbf{Average time of a call -- $t_M$}: This is the assumed average time of the calls of honest users. For the simulations, we assumed $t_M = 5$ time units; 
  
  \item \textbf{Dropping Factor} -- We assume the following values for the dropping factor function (Equation~\ref{eq:drop}):
  \begin{itemize}
    \item $p_\In = 2$;
    \item $p_\Wait = 8$;
    \item $\alpha = 1.89$.
  \end{itemize}
  These values were chosen so that the dropping factor increases in a reasonable fashion for calls with duration greater than $t_M$. Sample values are shown below, recalling that $t_M = 5$ seconds:
\bigskip
\begin{center}
\begin{tabular}{cc}
   \toprule
   Call Duration (mins) & Dropping Factor\\
   \midrule
   6 & 12.37\\
   8 & 17.31\\
   10 & 27.84\\
\bottomrule
\end{tabular}  
\end{center}
  \bigskip
That is, dropping factor of a call with duration of 10 minutes, \ie, $2 \times t_M$, is approximately 3 times greater than the dropping factor of a call whose status is $\wait$ (27.84/8). This is a reasonable ratio. However, according to the specific application other values can be set for $p_\In,p_\Wait$ and $\alpha$. Finally, the choice of setting $p_\Wait = 4 \times p_\In$ was selected so that the calls with duration less than $t_M$ have much less chance of being dropped than the calls that are still waiting for the responder.

    \item \textbf{Size of Buffer -- $k$}: This is the upper-bound on the size of the server's buffer denoting the processing capacity of the application. $k = 24$;
    \item \textbf{Rate of Calls ($R$)}: We fixed the total rate of calls to be $R$ which is the result of summing the rate of legitimate calls, $R_L$, and the rate of attacker calls, $R_A$. That is, $R_L + R_A = R$. 

    The value of $R$ is computed using standard techniques\footnote{Using the Erlang model which computes $R$ by  taking into account $k$ and $t_M$.} so that if $R_L = R$, \ie, the server is not under attack, then the server will not be overloaded. 

    With $R$ fixed, we set $R_L$ and $R_A$ to be $R_L + R_A = R$, but considered scenarios with different proportions for $R_A$ and $R_L$. This reflects the fact that \coordinatedcall attack uses low traffic and therefore, can bypass usual defenses based on network traffic analysis which normally monitor the total rate of calls $R$. 

    We considered 5 different proportions for $R_L$ and $R_A$ expressed in the percentage of the total number of calls $R$:
\bigskip
\begin{center}
\begin{tabular}{ccc}
   \toprule
   Legitimate Calls ($R_L$) & Attacker Calls ($R_A$)\\
   \midrule
   83\% & 17\%\\
   67\% & 33\%\\
   50\% & 50\%\\
   33\% & 67\%\\
   17\% & 83\%\\
\bottomrule
\end{tabular}  
\end{center}
  \bigskip
     \item \textbf{Total time of the simulation -- $total$}: This is the total time of the simulation using PVeStA. We 
    used in our simulations $total$ equal to 40 time units, similar to the time used in \cite{eckhardt12fase};
     \item \textbf{Delay of the Network}: We also assumed a delay of 0.1 time units of message in the network;
     \item \textbf{Degree of confidence for the simulation}: Our simulations were carried out with a degree of  confidence of 99\% (see \cite{agha06entcs,sen05cav} for more details on statistical model checking).
\end{itemize}

\paragraph{Quality Measures}
In our simulation, we use quality measures specific for VoIP services. 
These are specified by expressions of the QuaTEx quantitative, probabilistic temporal logic defined in~\cite{agha06entcs}.
We perform statistical model checking of our defense in the sense of~\cite{sen05cav}:  
once a QuaTEx formula and desired degree of confidence are specified, a sufficiently large number of Monte Carlo simulations are carried out allowing for the verification of the QuaTEx formula. The Monte Carlo simulations are carried out by the computational tool Maude~\cite{clavel-etal-07maudebook} and the statistical model checking is carried out by PVeStA.

The QuaTEx formulas, \ie, the quality measures, that we use in our simulations are defined below.
The operator $\bigcirc$ is a temporal modality that specifies the advancement of the global time to the time of
the next event (see~\cite{agha06entcs} for more details).

\begin{itemize}
	\item Complete: How many honest calls were able to stay in the $\incall$ status for the expected duration.

		 \begin{center}
		 \begin{tabular}{ll}
		 $complete(total)$ = & if $time > total$ then  $\frac{countComplete}{countHonest}$  
     \\[5pt]& \qquad
     else $\bigcirc\ complete(total)$ \\[5pt]
		 \end{tabular}  
		 \end{center}
		where $countComplete$ is a counter that is incremented whenever an honest call is completed.
		
	\item Incomplete: How many honest calls were able to have the $\incall$ status but were dropped before completing the call, \ie\ not staying in $\incall$ status for the expected duration;
	
	 \begin{center}
		 \begin{tabular}{ll}
		 $incomplete(total)$ = & if $time > total$ then  $\frac{countIncomplete}{countHonest}$  
		 \\[5pt]& \qquad
		 else $\bigcirc\ incomplete(total)$ 
		 \end{tabular}  
		 \end{center}
		where $countIncomplete = countIncall - countComplete $ and $countIncall$ is a counter that is incremented whenever an honest calls changes from status $\wait$ to $\incall$.
	
	\item Unsuccessful: How many honest calls were not even able to reach the $\incall$ status. That is, how many calls were not even able to start talking between each other.
	
	 \begin{center}
		 \begin{tabular}{ll}
		 $unsuccessful(total)$ = & if $time > total$ then  $\frac{countUnsuccessful}{countHonest}$  
		 \\[5pt]& \qquad
		 else $\bigcirc\ unsuccessful(total)$ \\[5pt]
		 \end{tabular}  
		 \end{center}
		where $countUnsuccessful = countHonest - countIncall$.
		
	\item The average of client incomplete calls: We also measure the average proportion of time  legitimate clients were able to talk in an incomplete call before they were dropped.
		 \begin{center}
		 \begin{tabular}{ll}
		   $\mathit{avgInCall}(total)$ = & if $time > total$ then  $\frac{totalTimeInCall}{totalIncompleteCall}$  
		 \\[5pt]& \qquad
		 else $\bigcirc\ \mathit{avgInCall}(total)$ \\[5pt] 
		 \end{tabular}  
		 \end{center}

		where $totalTimeInCall$ is the sum of how much percent of time clients were able to talk before being interrupted and the $totalIncompleteCall$ is the total of clients that were not able to finish their call.
			
\end{itemize}


We carried out simulations with the three different types of dropping strategies described in Section~\ref{sec:seven}, namely uniform, roulette and $\frac{k}{2}$-tournament. We also carried out simulations with a scenario without \seven.

\subsection{No Defense}

\begin{figure}
\subfigure[Exponential Call Duration.]{
  \centering
\includegraphics[width=0.55\textwidth]{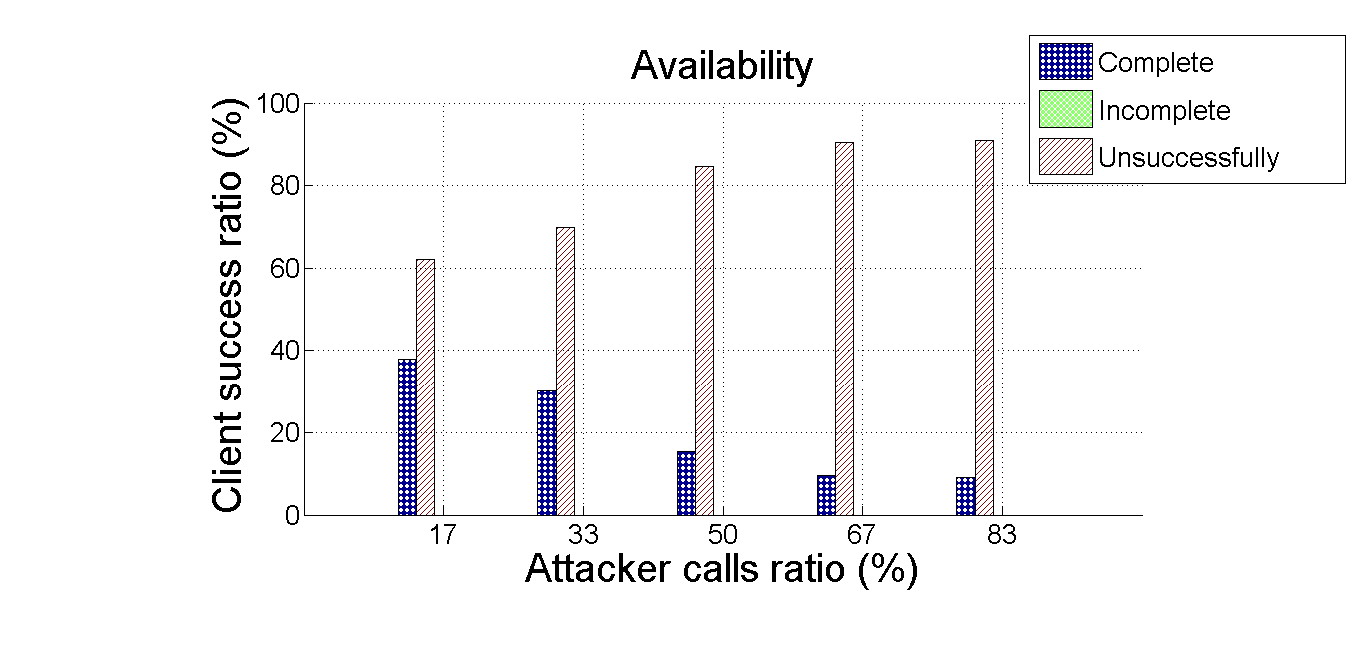}
\label{fig:exp_no_def_sim}
}
\hspace{-21.5mm}
\subfigure[Lognormal Call Duration.]{
  \centering
\includegraphics[width=0.55\textwidth]{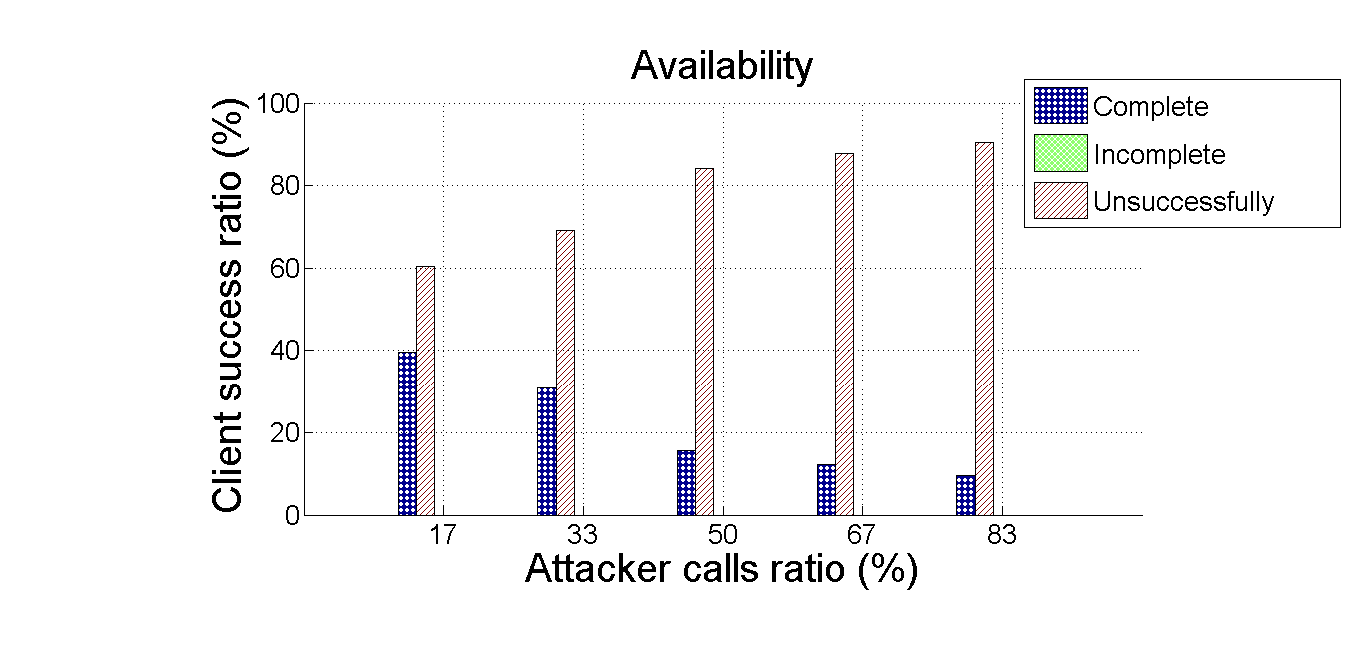}
\label{fig:log_no_def_sim}
}
\caption{Client Success Ratio: Simulation Results when not using \seven.}
\label{fig:dip_no_defense_sim}
\end{figure}

Our simulations results are depicted in Figure~\ref{fig:dip_no_defense_sim}. They suggest that the \coordinatedcall attack is indeed effective in reducing the availability of a VoIP service when assuming both an exponential and a lognormal call duration. Increasing the proportion of attackers rapidly increases the proportion of Unsuccessful calls, \ie, calls that did not even start a conversation, while the proportion of Complete calls falls. As expected there are no Incomplete calls as the VoIP server does not interrupt calls.

\subsection{Uniform Dropping Strategy}

\begin{figure}
\subfigure[Exponential Call Duration.]{
  \centering
\includegraphics[width=0.55\textwidth]{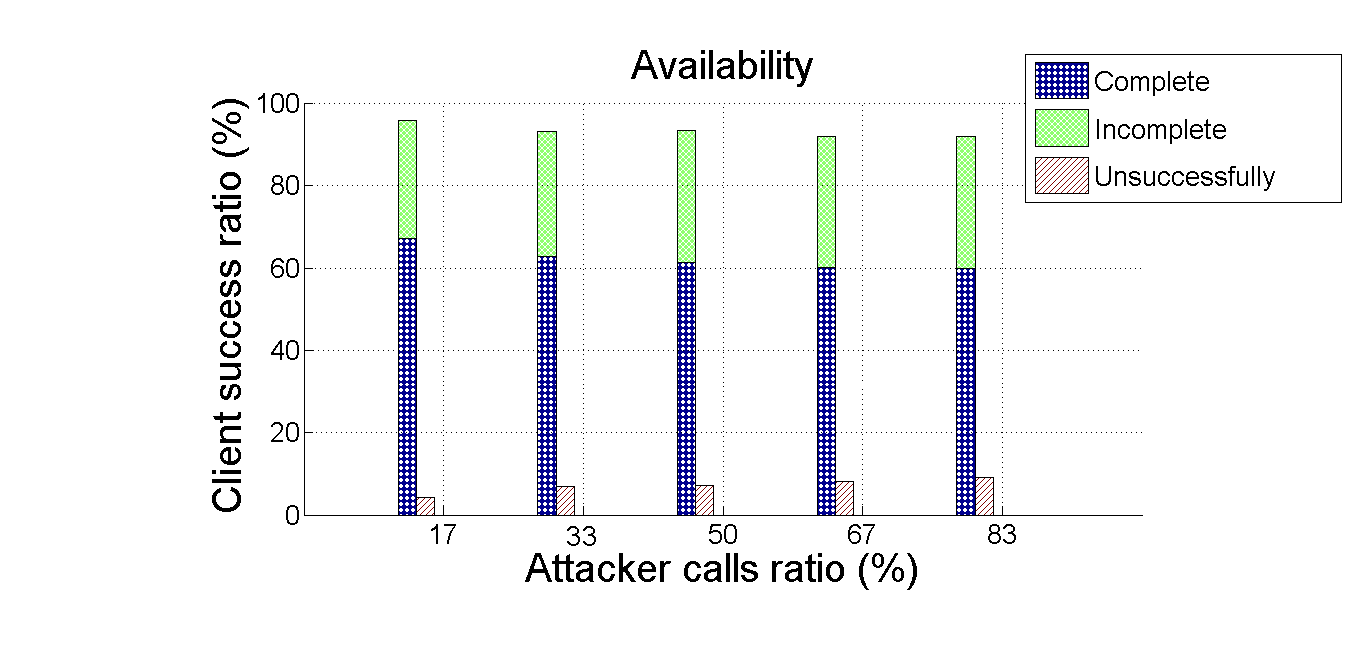}
\label{fig:exp_uni_sim}
}
\hspace{-21.5mm}
\subfigure[Lognormal Call Duration.]{
  \centering
\includegraphics[width=0.55\textwidth]{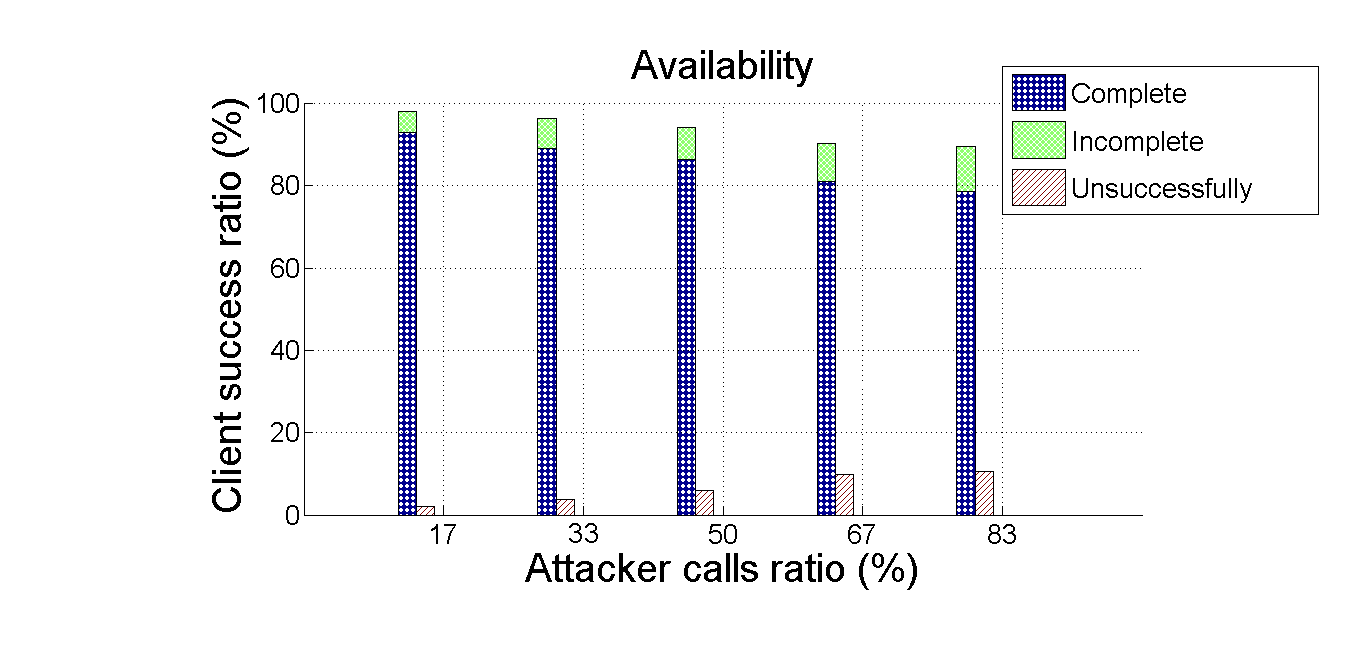}
\label{fig:log_uni_sim}
}
\subfigure[Exponential Call Duration.]{
  \centering
\includegraphics[width=0.55\textwidth]{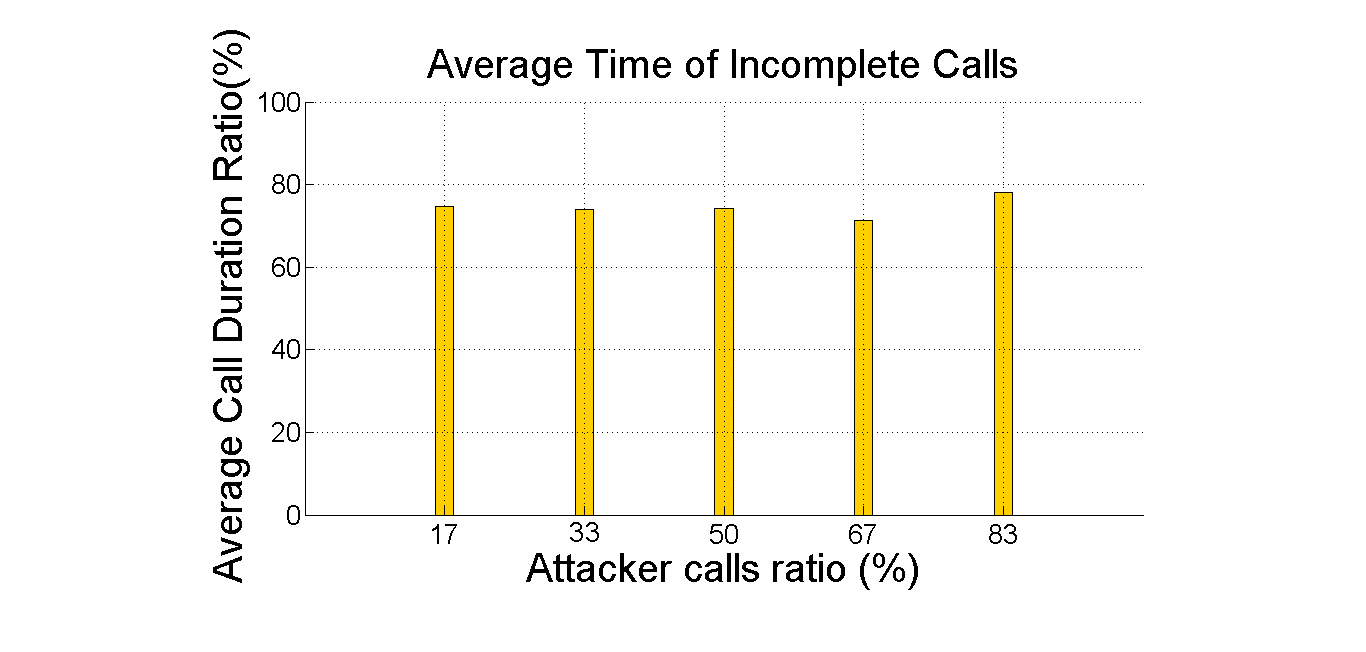}
\label{fig:exp_uni_inc_sim}
}
\hspace{-12mm}
\subfigure[Lognormal Call Duration.]{
  \centering
\includegraphics[width=0.55\textwidth]{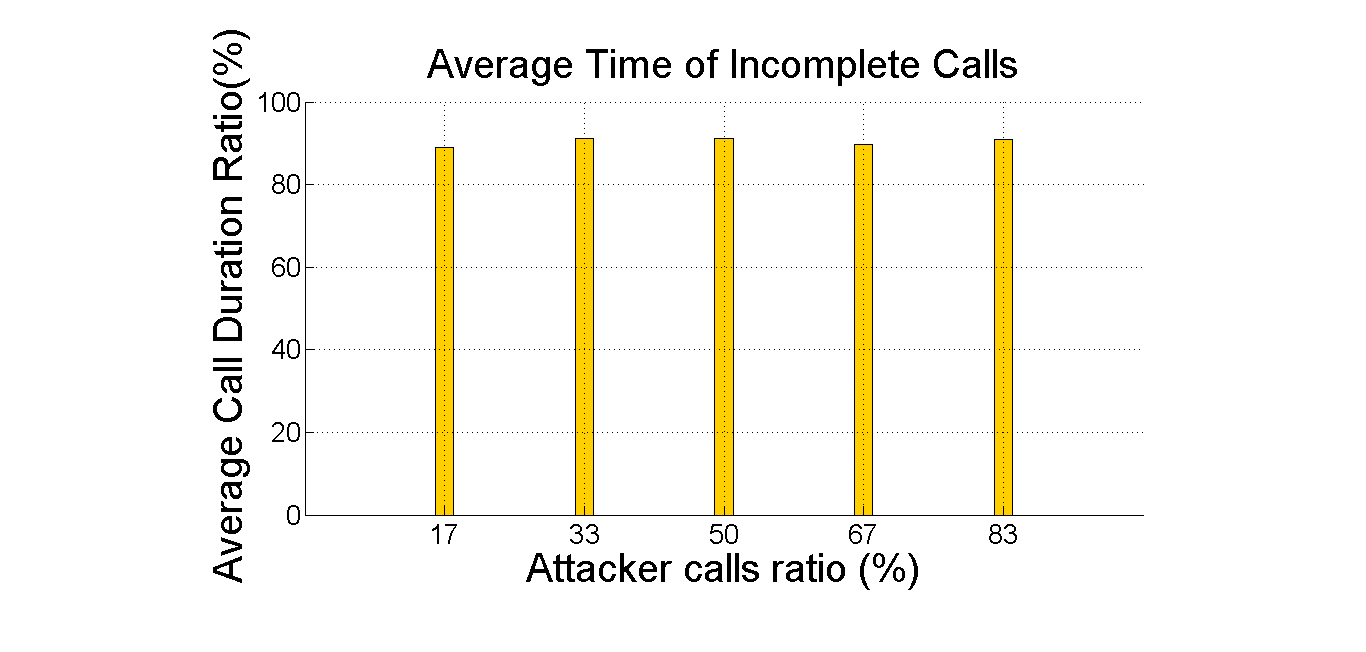}
\label{fig:log_uni_inc_sim}
}
\caption{Client Success Ratio and Average Time of Incomplete Calls: Simulation Results when a uniform dropping strategy.}
\label{fig:dip_uniform_sim}
\end{figure}
\FloatBarrier

Figure~\ref{fig:dip_uniform_sim} depicts the results when using \seven with a uniform dropping strategy. It suggests that \seven can indeed mitigate the \coordinatedcall attack. The proportion of Complete calls remains at high levels when assuming both an exponential, above 60\% of legitimate calls, and a lognormal call duration, above 80\%.  

As \seven may interrupt calls in the middle of a conversation, there are Incomplete calls, \ie, calls where the parties have started to communicate, but were interrupted before communicating for the expected time.
For exponential call duration, around 30\% of legitimate calls were interrupted, while for lognormal call duration around 10\% of legitimate calls were interrupted. However, the average time of incomplete calls suggests that although these calls are interrupted, they still are able to communicate for long periods of time, above 70\% of the expected time for exponential call duration and above 89\% of the expected time for lognormal call duration.

\subsection{Roulette Dropping Strategy}

\begin{figure}
\subfigure[Exponential Call Duration.]{
  \centering
\includegraphics[width=0.55\textwidth]{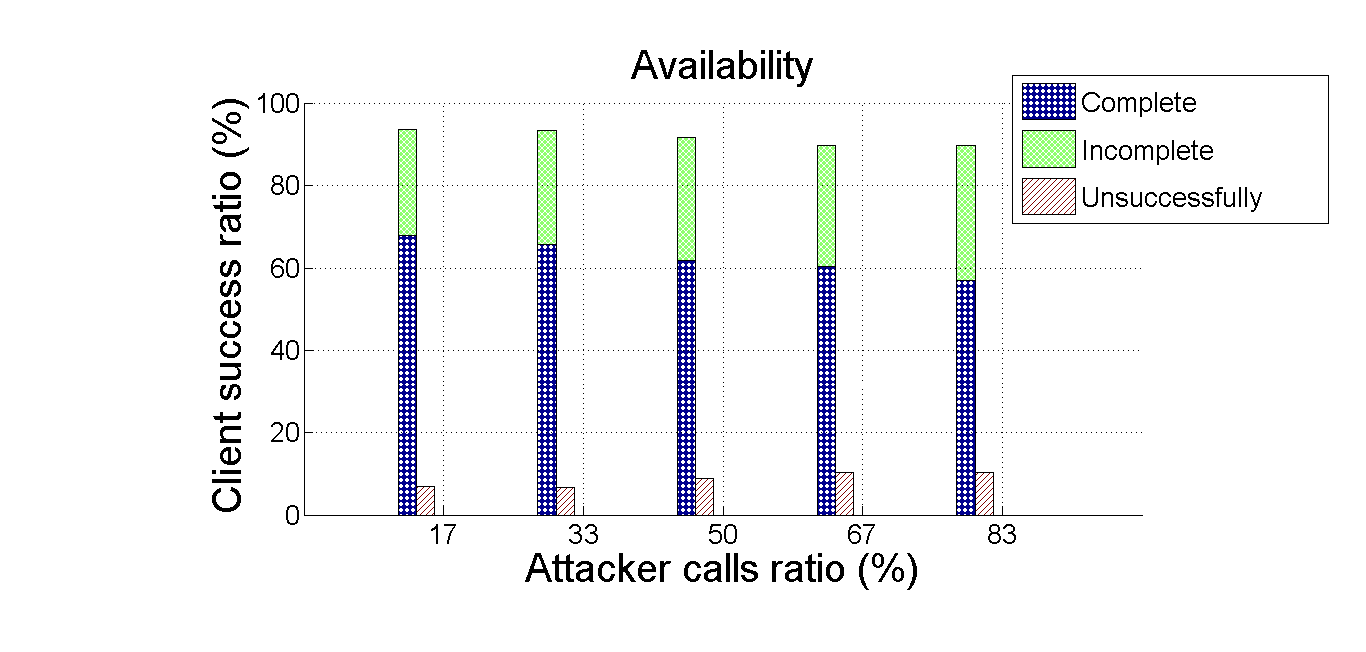}
\label{fig:exp_roulette_sim}
}
\hspace{-22mm}
\subfigure[Lognormal Call Duration.]{
  \centering
\includegraphics[width=0.55\textwidth]{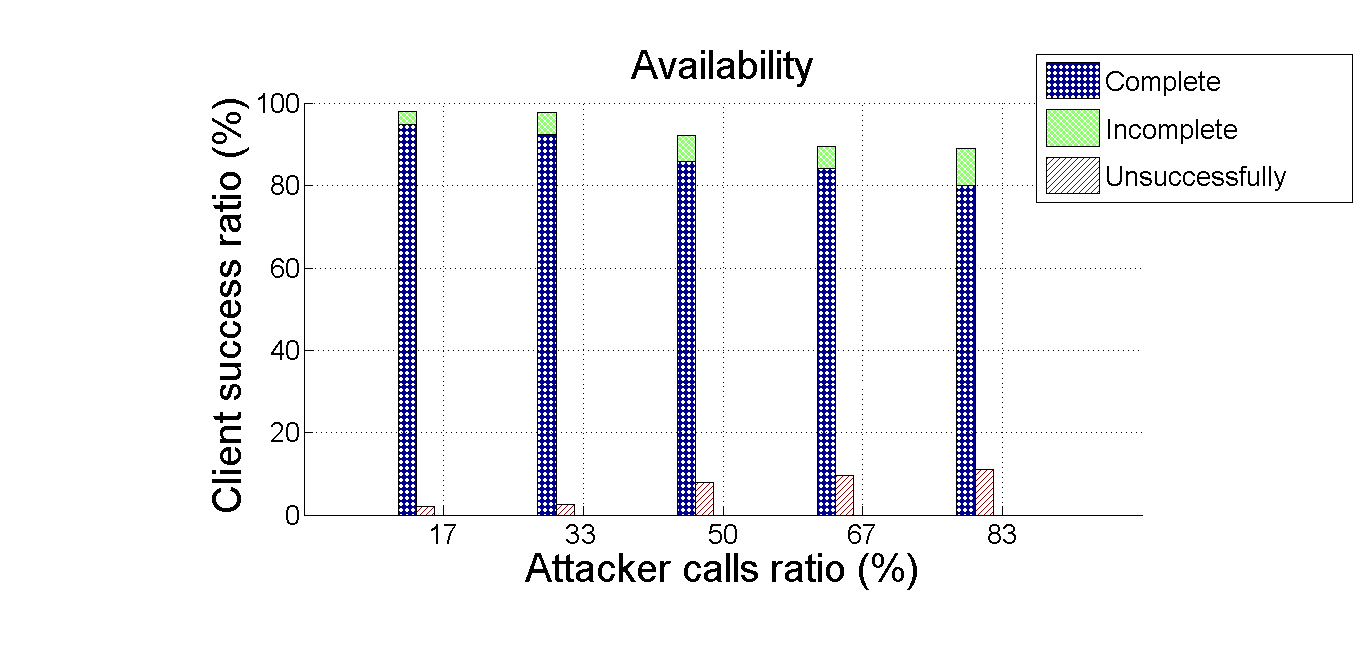}
\label{fig:log_roulette_sim}
}
\subfigure[Exponential Call Duration.]{
  \centering
\includegraphics[width=0.55\textwidth]{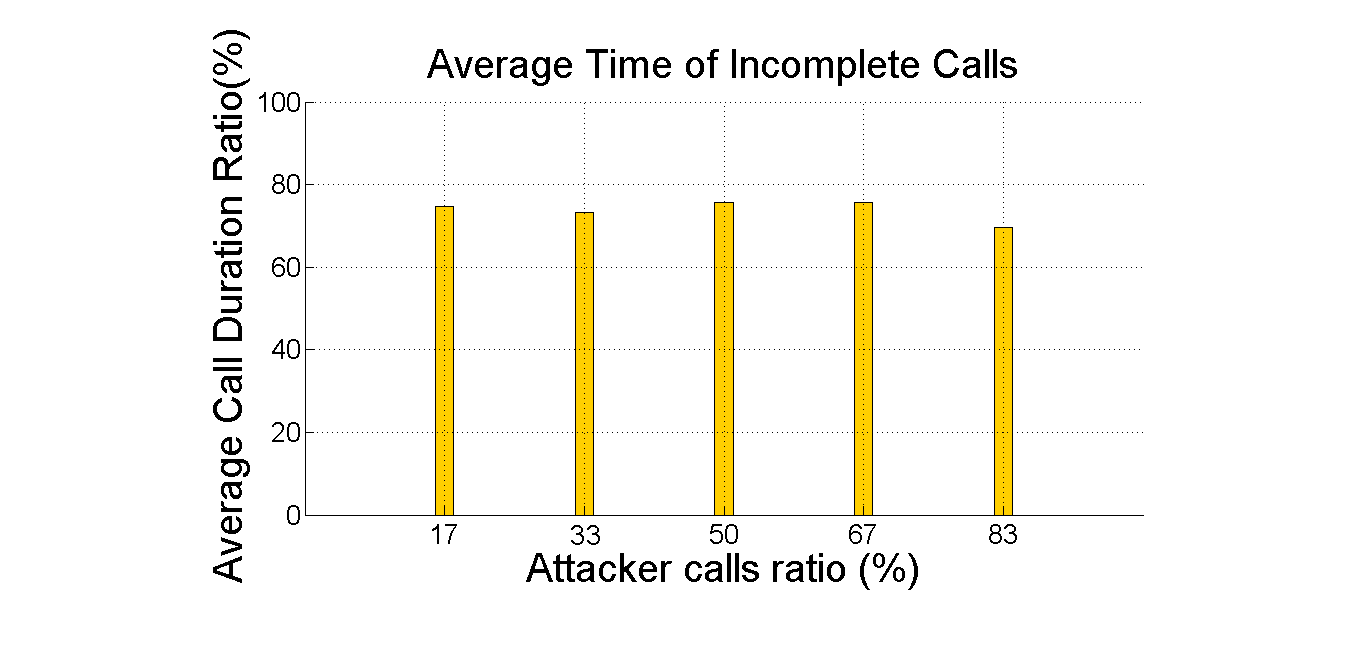}
\label{fig:exp_roulette_inc_sim}
}
\hspace{-12mm}
\subfigure[Lognormal Call Duration.]{
  \centering
\includegraphics[width=0.55\textwidth]{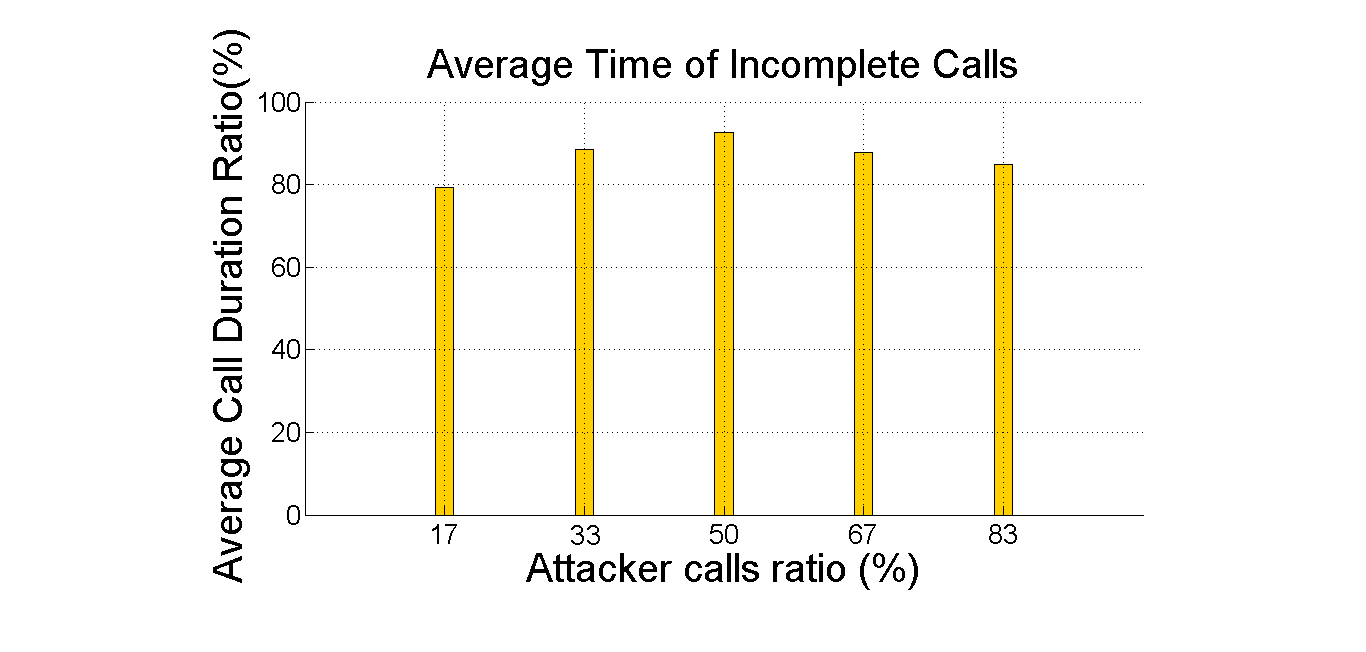}
\label{fig:log_roulette_inc_sim}
}
\caption{Client Success Ratio and Average Time of Incomplete Calls: Simulation Results when a roulette dropping strategy.}
\label{fig:dip_roulette_sim}
\end{figure}

Figures~\ref{fig:dip_roulette_sim} depicts our simulation results when using a roulette dropping strategy. The results are similar to the results obtained with the uniform dropping strategy. For exponential call duration, above 60\% of legitimate calls were completed and around 30\% were interrupted by \seven. For lognormal call duration, above 80\% of legitimate calls were completed and around 10\% were interrupted by \seven. Moreover, the interrupted calls stayed communicating in average above 70\% of the expected call duration for exponential call duration and above 88\% of the expected call duration for lognormal call duration. 

\subsection{$\frac{k}{2}$-tournament}

\begin{figure}
\subfigure[Exponential Call Duration.]{
  \centering
\includegraphics[width=0.55\textwidth]{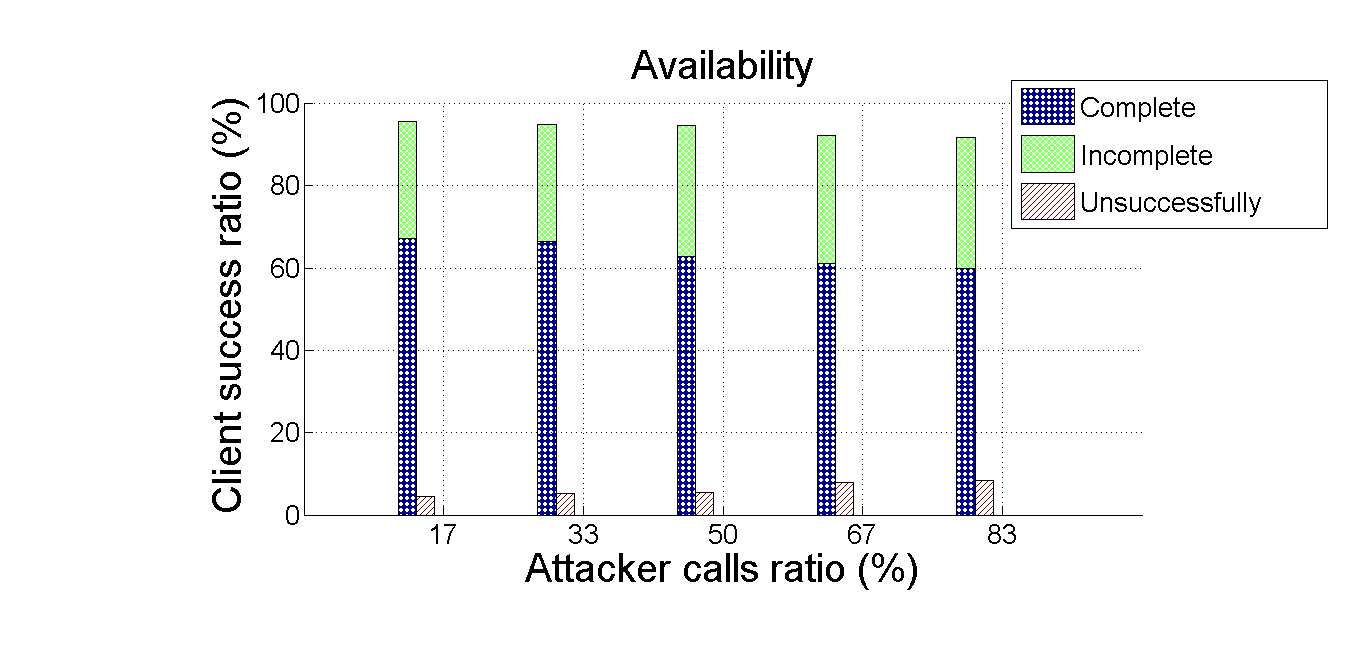}
\label{fig:exp_tour_sim}
}
\hspace{-22mm}
\subfigure[Lognormal Call Duration.]{
  \centering
\includegraphics[width=0.55\textwidth]{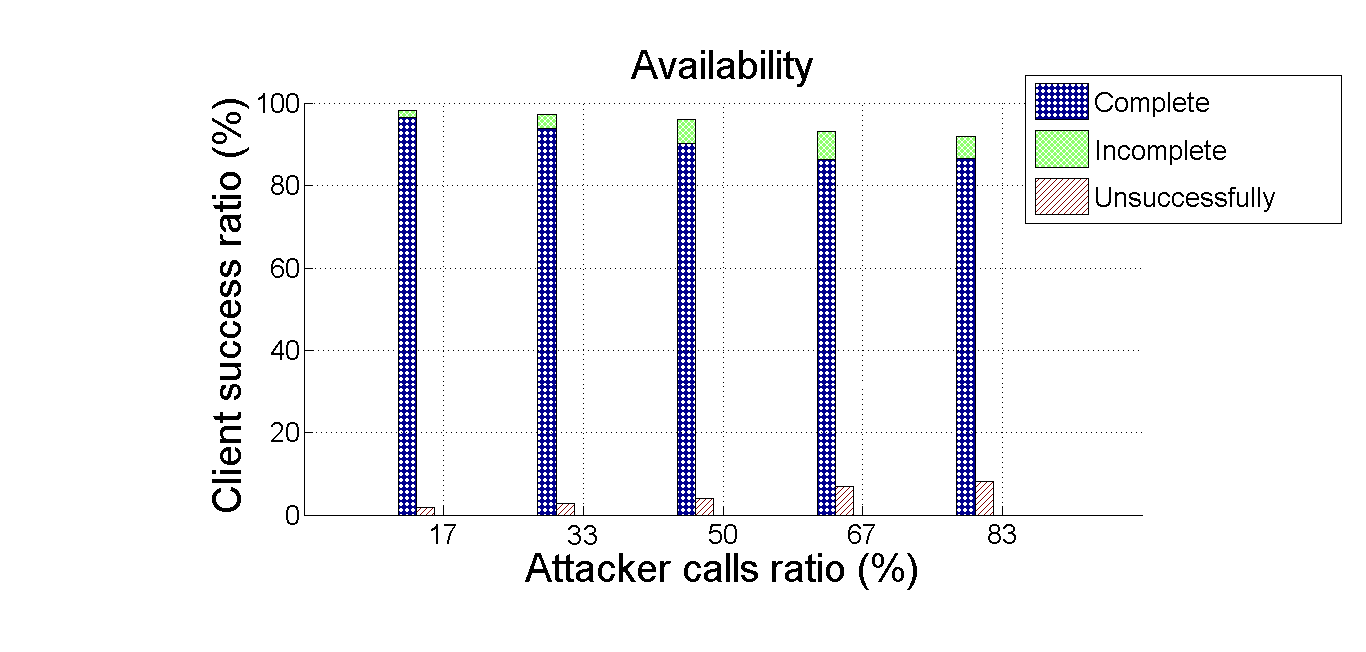}
\label{fig:log_tour_sim}
}
\subfigure[Exponential Call Duration.]{
  \centering
\includegraphics[width=0.55\textwidth]{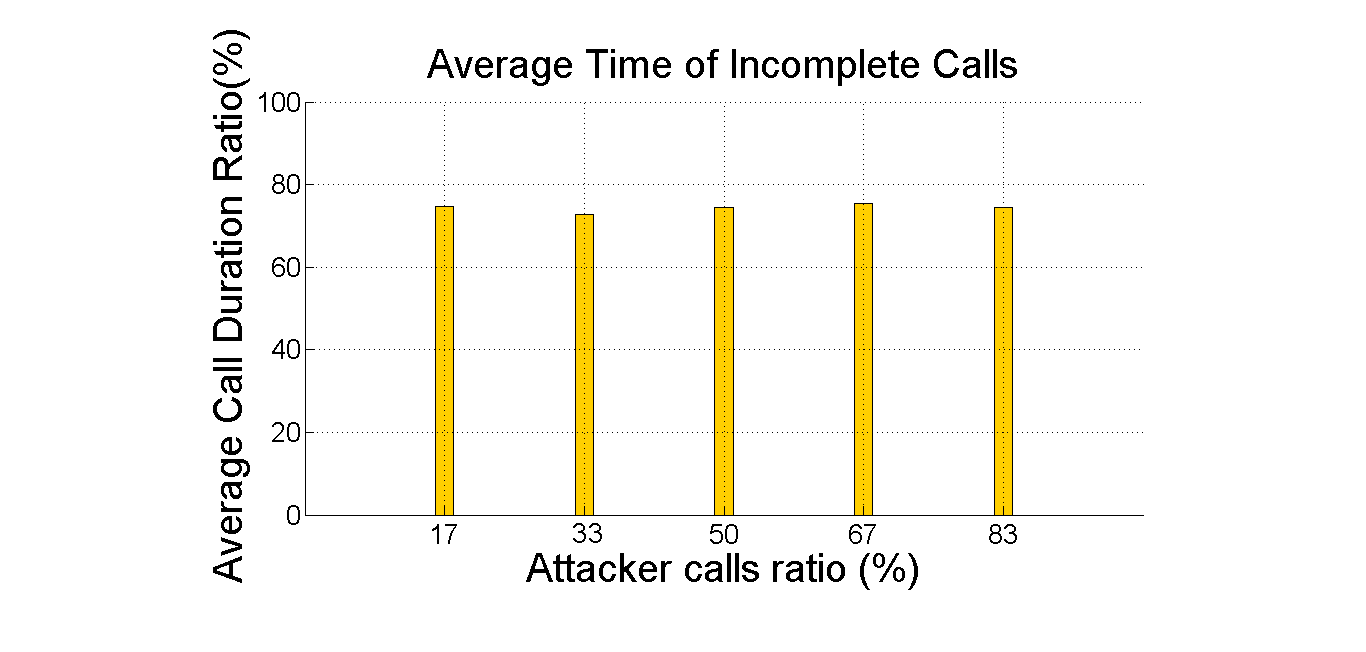}
\label{fig:exp_tour_inc_sim}
}
\hspace{-12mm}
\subfigure[Lognormal Call Duration.]{
  \centering
\includegraphics[width=0.55\textwidth]{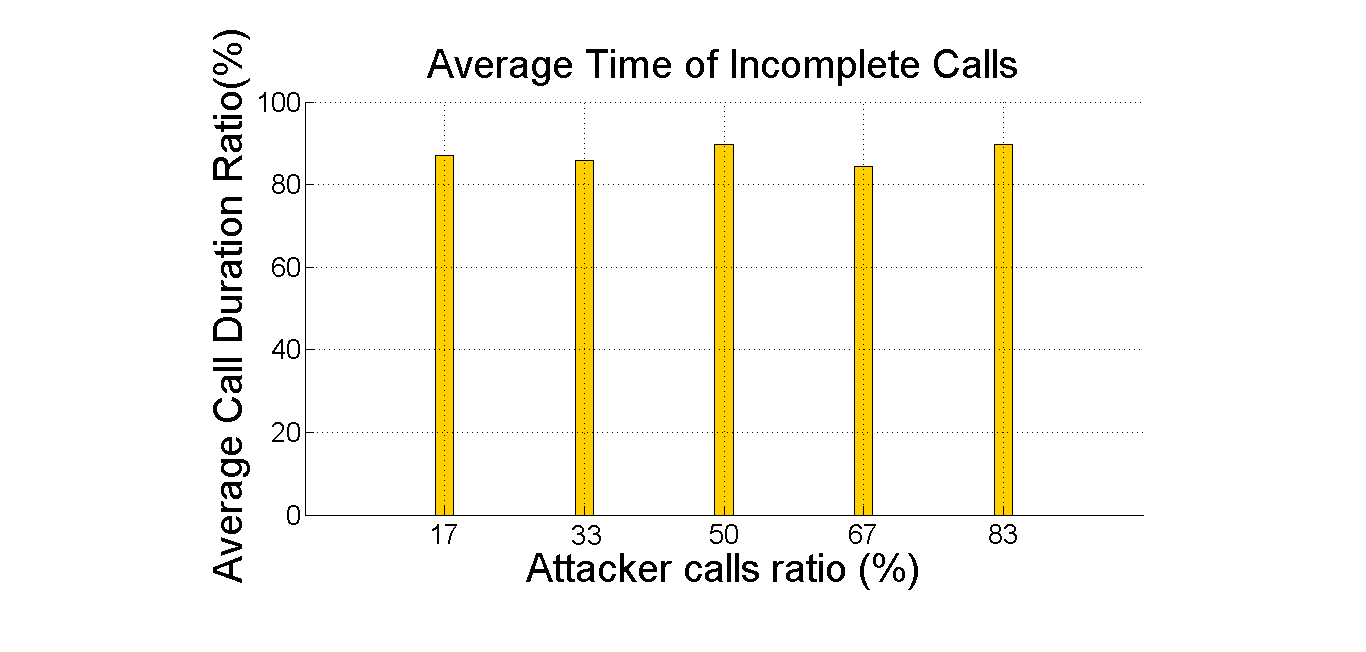}
\label{fig:log_tour_inc_sim}
}
\caption{Client Success Ratio and Average Time of Incomplete Calls: Simulation Results when using a roulette dropping strategy.}
\label{fig:dip_tour_sim}
\vspace{-3mm}
\end{figure}

Figures~\ref{fig:dip_tour_sim} and \ref{fig:tour_inc_sim} depict the simulation results obtained by using a $\frac{k}{2}$-tournament dropping strategy. They suggest that this strategy is better than the roulette and uniform strategies. The proportion of complete calls is above 62\% for exponential call duration and above 86\% for lognormal call duration. In other words, around 30\% of legitimate calls were interrupted by \seven when assuming exponential call duration and around 10\% of legitimate calls were interrupted by \seven when assuming lognormal call duration. Moreover, the average time of incomplete calls is always greater than 72\% for exponential call duration and above 84\% for lognormal call duration.

\FloatBarrier

%% file: structure/exp.tex

In our experiments, we used Asterisk version 13.6.0 which is a SIP server widely used by small and mid size companies for implementing their VoIP services. We assume there are honest users and malicious attackers which try to make the VoIP unavailable. Both the traffic of the honest users and the attackers are emulated using the tool SIPp~\cite{chico:bioca} version 3.4.1. SIPp generates calls which may be configured as the caller or the callee. Thus, in our experiments, we used pairs of SIPp, one pair for generating the honest user calls and the other pair for generating the attacker calls. Finally, we developed the \seven proxy in C++ which implements the selective strategy described in Section~\ref{sec:seven}.

\begin{center}
  \includegraphics[width=6cm]{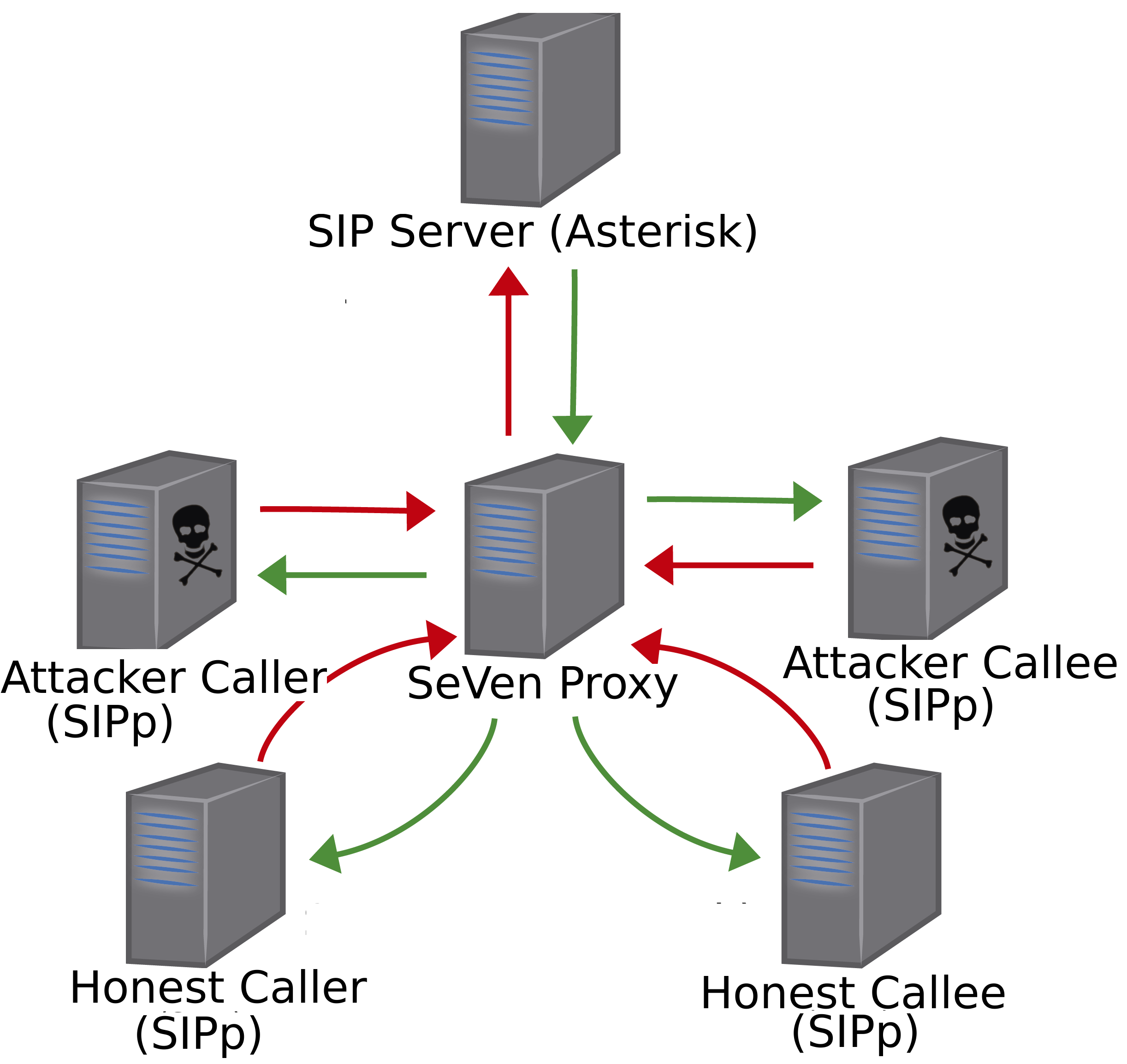}
\end{center}
  
The figure above illustrates the topology of the experiments we carried out. To make a call, the pairs of SIPp send messages to the \seven\ proxy which on the other hand forwards them to Asterisk. Similarly, any message generated by Asterisk is forward to the \seven\ proxy which then forwards them to the corresponding users. Therefore, \seven\ is acting as an \emph{Outbound Proxy} for both Asterisk and the pairs of SIPp. {For our experiments, it is enough to use a single machine. We used a machine with configuration Intel(R) Core(TM) i7-4510U CPU @ 2.00GHz and 8 GB of RAM.}

\paragraph{Parameters} 
We use the following parameters to configure our experiments:

\begin{itemize}
  \item \textbf{Average Call Duration ($t_M$)} -- We assume known what is the average duration of calls. This can be determined in practice by analyzing the history of calls. We assume in our experiments that $t_M = 160$  seconds (approximately 2.6 minutes);

\item \textbf{SIP Sever Capacity ($k$)} -- This is the number of simultaneous calls the SIP server can handle. We set $k = 200$ which is a realistic capacity for a small company allowing 400 users ($2\times 200$) to use the service at the same time.

\item \textbf{Experiment Total Time ($T$)} -- Each one of our experiments had a duration of $60$ minutes which corresponds to 3600 calls in each experiment. With this duration, it was already possible to witness the damage caused by the Coordinated Call Attack as well as the efficiency of our solution for mitigating this attack.

\item \textbf{Traffic Rate ($R$)} -- Using a server with capacity of $k$, we calculated using standard techniques~\cite{jewett80book} what would be a typical traffic of such a server. It is $R = 60$ calls per minute. This value is computed using traditional techniques~\cite{jewett80book} (Erlang model) taking into account $k = 200$ and $t_M = 160$. Thus the service can handle $R$ legitimate calls per second. However, as the attacker does not behave as legitimate placing calls with much greater durations, the server can be subject to this attack. 

In our experiments, we split this rate among clients and attackers. This is because we want to emulate the fact that \coordinatedcall attack can deny service using low traffic thus bypassing usual defenses~\cite{Ha:2009:DIS:1655925.1656137,5426267,6195475,6720187,5541813,5630386} based on network traffic analysis or monitoring the number of incoming calls. This means that the total traffic (attacker + client) is always less or equal than $R$. 
\end{itemize}

The following graph illustrates client usage in normal conditions, \ie, without suffering an attack, using the parameters as specified above.
  \begin{center}
  \includegraphics[width=8cm]{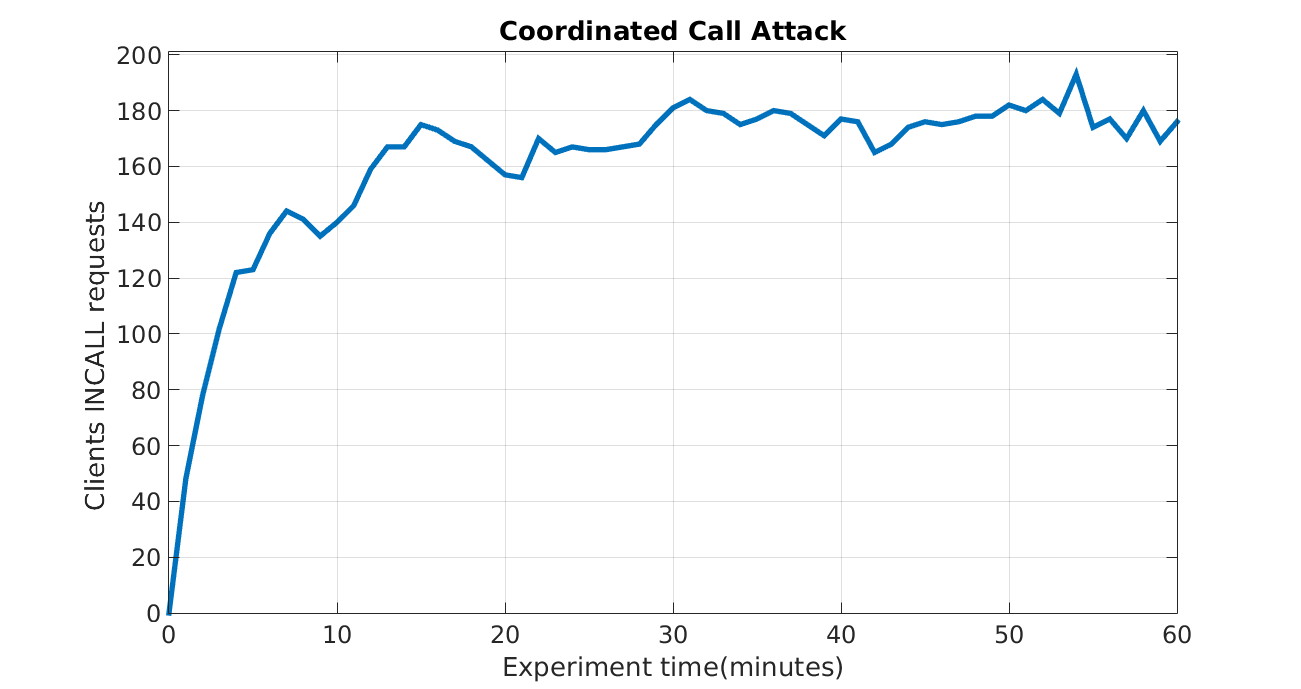}
  \end{center}
It shows that the server is indeed well dimensioned for this rate of (legitimate) calls. Clients occupy in average approximately 85\% of Asterisk's capacity thus not overloading the server, but still being heavily used.

Finally, we set the duration of the calls generated by SIPp as follows:
\begin{itemize}
  \item \textbf{Total Call Duration of Clients:} As described in Section~\ref{sec:seven}, we used two models for the call duration of legitimate client calls, namely the exponential model suitable for non-VoIP calls and the lognormal model suitable for VoIP calls. The parameters, $\lambda, \mu,\sigma$, in Equations~\ref{eq:exp} and \ref{eq:lognormal} were computed as described in~\cite{Brown_2002a,7414101} using the average call duration $t_M$.
Whenever we generate a new call, we generate the call duration randomly according to the used model (exponential or lognormal). SIPp ends the call when its corresponding call duration is reached.
  \item \textbf{Call Duration of Attackers:} Following the Coordinated Call Attack, we do not limit the call duration of an attacker call. His calls communicate for indefinite time. 
\end{itemize}

\paragraph{Quality Measures} For our experiments, we used the following three  quality measures for our calls:
\begin{itemize}
  \item \textbf{Complete Call:} A call is complete whenever its status changed from $\wait$ to $\incall$ and \emph{it is able to stay in status $\incall$ for its corresponding call duration}. That is, the caller was able to communicate with the responder for all the prescribed duration;
  \item \textbf{Incomplete Call:} A call is incomplete whenever its status changed from $\wait$ to $\incall$, but \emph{it was not able to stay in status $\incall$ for its corresponding call duration}. That is, the caller was interrupted before completing the call;
  \item \textbf{Unsuccessful Call:} A call is unsuccessful if it did not even change its status from $\wait$ to $\incall$. That is, the caller did not even have the chance to speak with the responder.
\end{itemize}

Intuitively, complete calls are better than incomplete calls which are better than unsuccessful calls. In order to support this claim, we also computed the average duration call of the incomplete calls, that is, the time that users in average were able to stay communicating before they were interrupted by \seven. 

\subsection{Experimental Results}

We carried out the corresponding experiments to the scenarios used in Section~\ref{sec:sim}. That is, we tested the efficiency of the \coordinatedcall attack when the server is not running any defense. We also carried out experiments with scenarios using an exponential and lognormal call duration with the uniform, roulette and 100-tournament dropping strategies. 

\subsubsection{No Defense}
\begin{figure}
\subfigure[Exponential Call Duration.]{
  \centering
\includegraphics[width=0.55\textwidth]{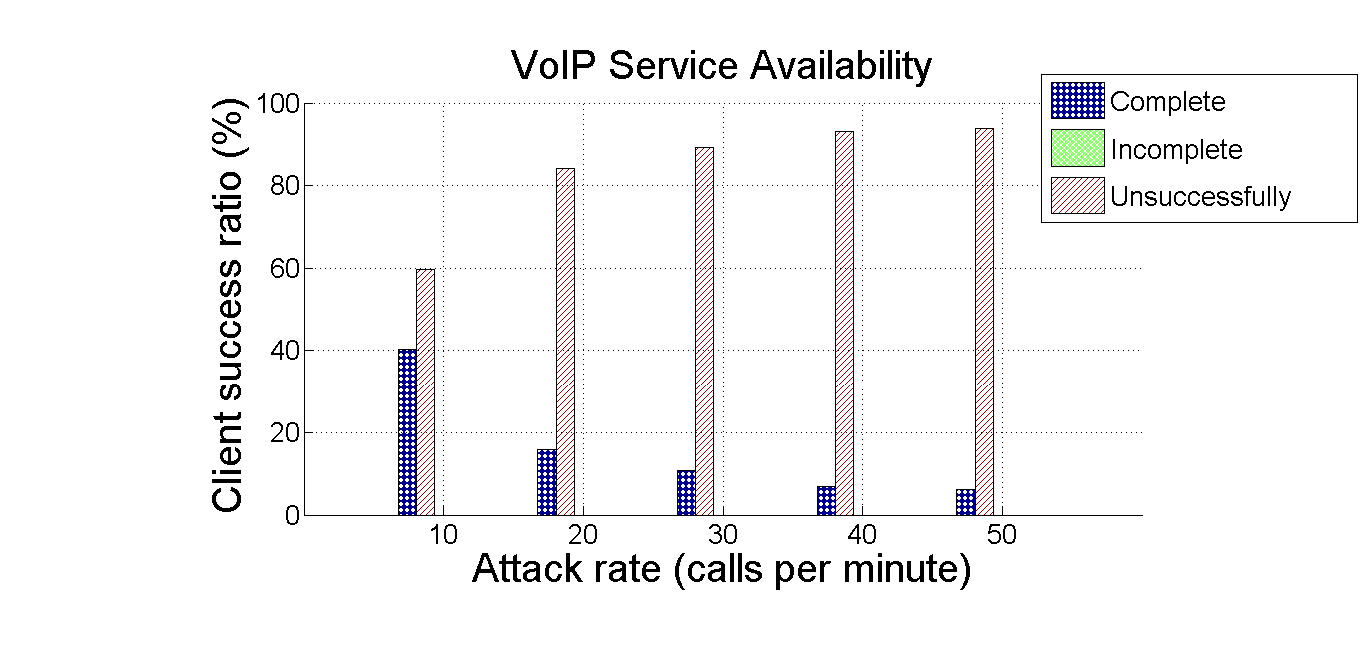}
\label{fig:exp_no_def}
}
\hspace{-22mm}
\subfigure[Lognormal Call Duration.]{
  \centering
\includegraphics[width=0.55\textwidth]{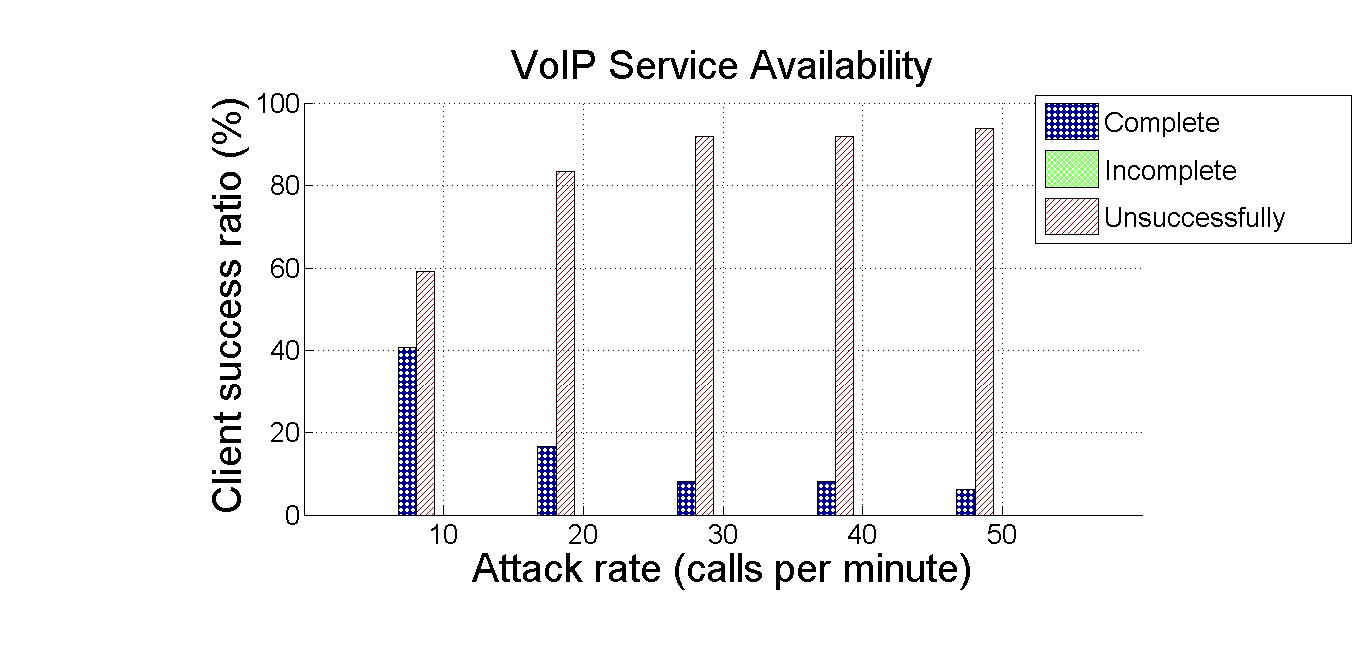}
\label{fig:log_no_def}
}
\caption{Client Success Ratio: Experimental Results when not using \seven.}
\label{fig:no-defense-disp}
\end{figure}

\begin{figure}[t]
\begin{center}
\includegraphics[width=0.66\textwidth]{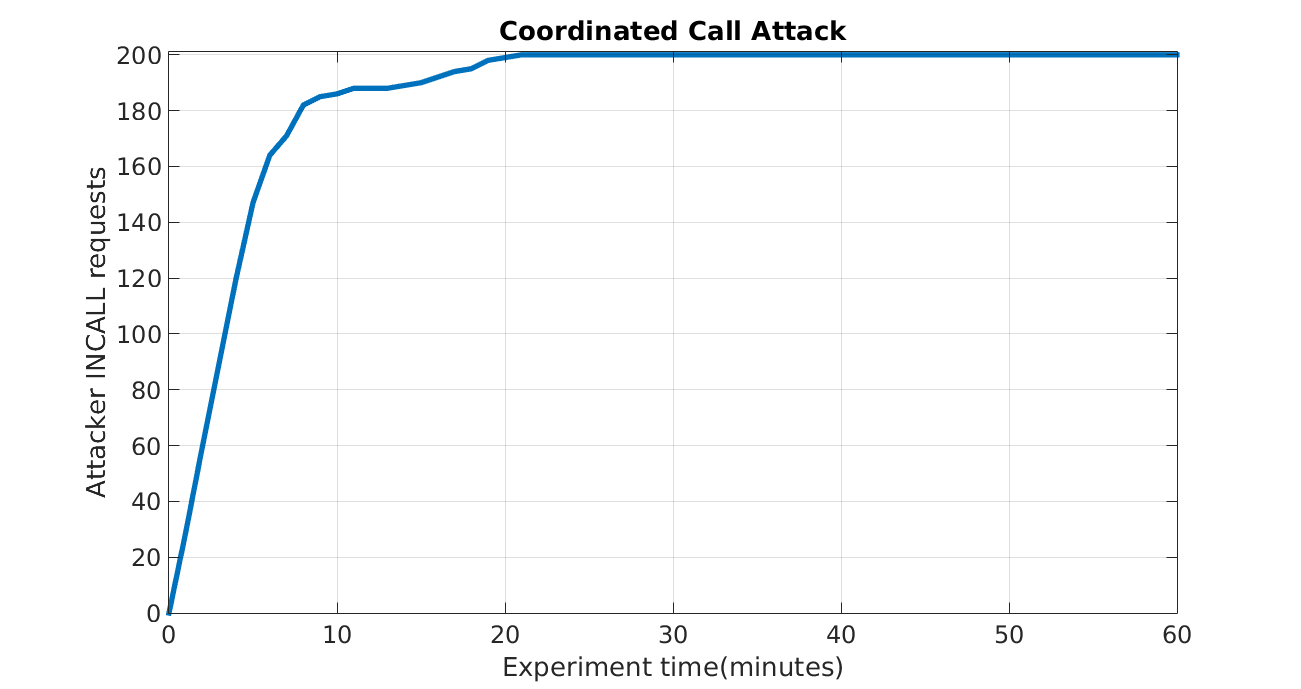}
\end{center}
\vspace{-2mm}
\caption{Attacker Call Occupancy in Buffer: Experimental Results when not using \seven.}
\label{fig:no_def_occ}
\vspace{-2mm}
\end{figure}

Figures~\ref{fig:no-defense-disp} and~\ref{fig:no_def_occ} illustrate our main results when assuming exponential and lognormal call duration and not using any defense mechanism. Our results demonstrate the efficiency of the \coordinatedcall attack. We observed that the VoIP availability decreases considerably when increasing the proportion of attacker calls in the rate $R$ (Figure~\ref{fig:no-defense-disp}). In particular, the number of unsuccessful call increases to level near 100\%, while the number of completed calls falls to close to 0\%. Moreover, there are no incomplete calls, which is expected since no calls are interrupted.

We also measured the number of attacker calls that the server serves during the experiment (Figure~\ref{fig:no_def_occ}). As expected from the profile of the \coordinatedcall attack, the attacker is able to deny service by slowly (after 10 minutes) occupying all the available calls in the server and therefore deny its service to legitimate clients.

\FloatBarrier

\subsubsection{Uniform Defense}

\begin{figure}[h]
\subfigure[Exponential Call Duration.]{
  \centering
\includegraphics[width=0.55\textwidth]{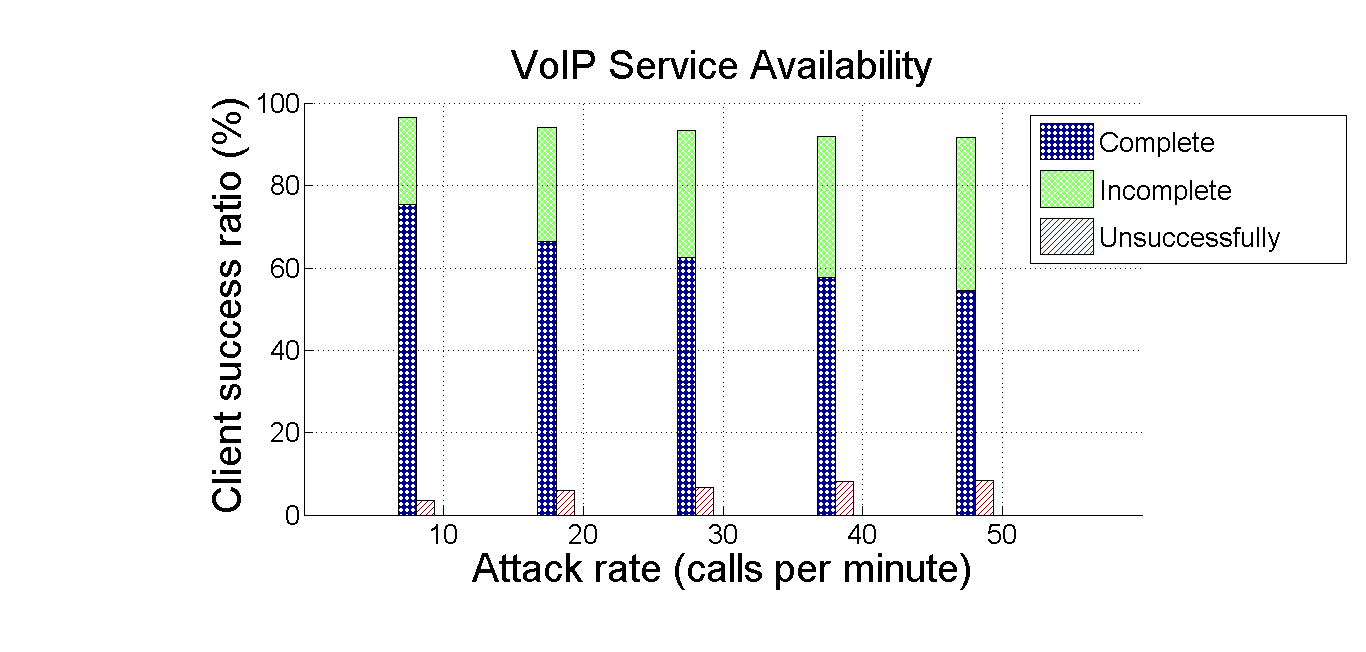}
\label{fig:exp_uni}
}
\hspace{-22mm}
\subfigure[Lognormal Call Duration.]{
  \centering
\includegraphics[width=0.55\textwidth]{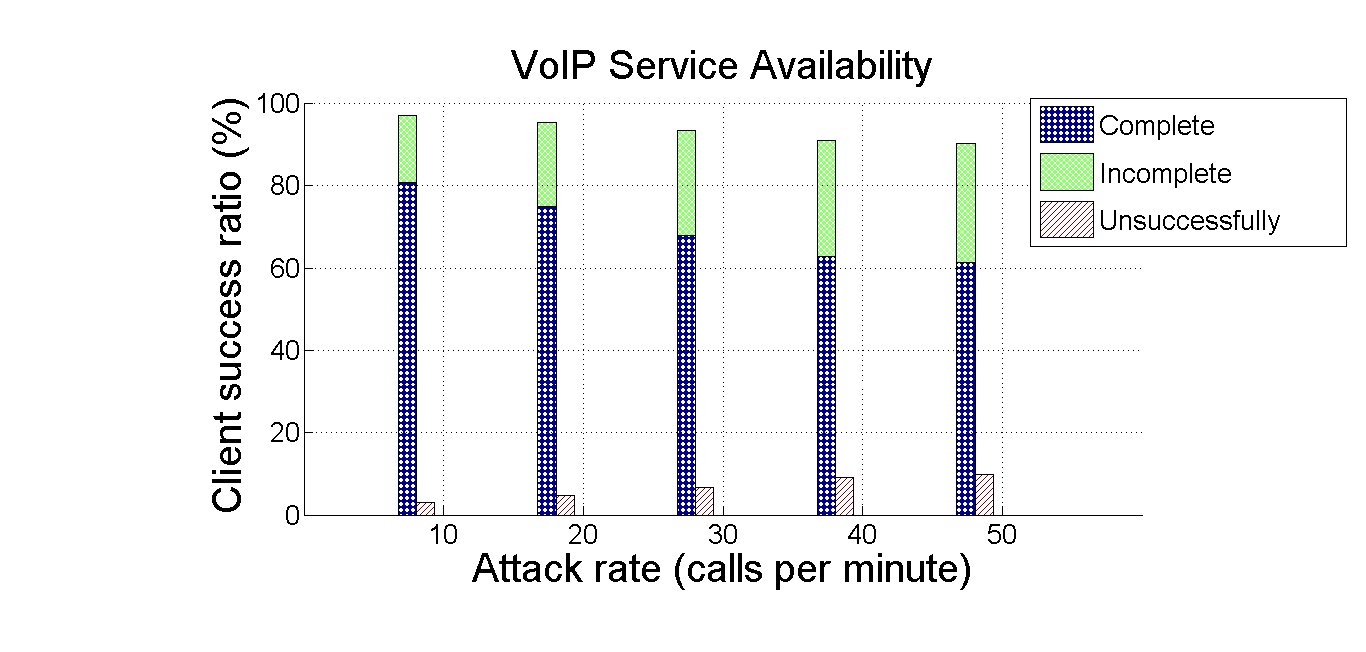}
\label{fig:log_uni}
}
\subfigure[Exponential Call Duration.]{
  \centering
\includegraphics[width=0.55\textwidth]{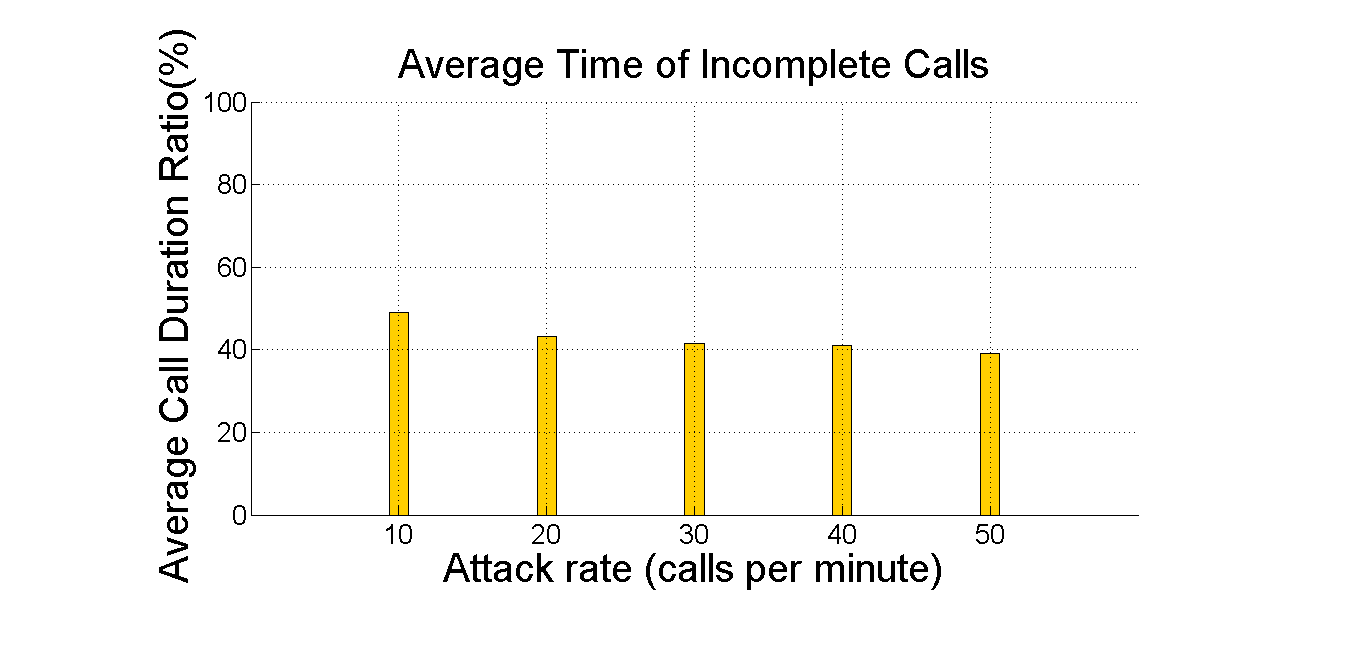}
\label{fig:exp_uni_inc}
}
\hspace{-12mm}
\subfigure[Lognormal Call Duration.]{
  \centering
\includegraphics[width=0.55\textwidth]{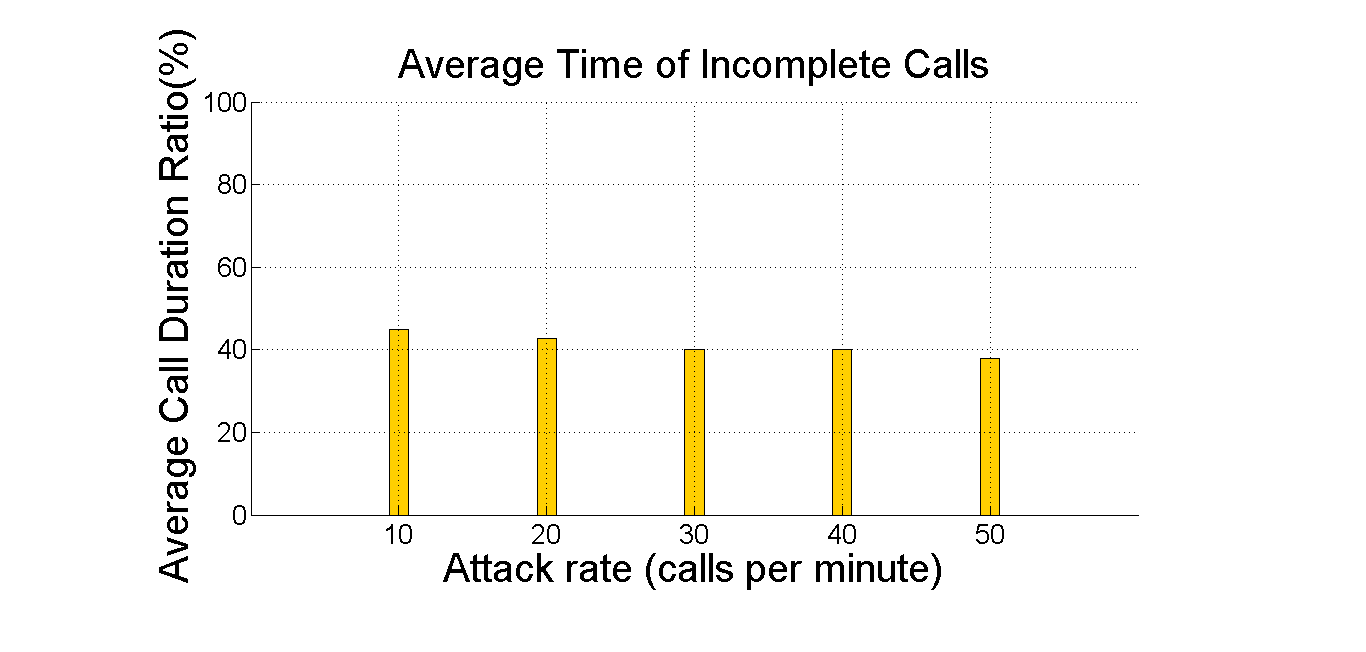}
\label{fig:log_uni_inc}
}
\caption{Client Success Ratio and Average Time of Incomplete Calls: Experimental Results when \seven and a uniform dropping strategy.}
\label{fig:uni-disp}
\end{figure}

\begin{figure}[t]
\begin{center}
\includegraphics[width=0.66\textwidth]{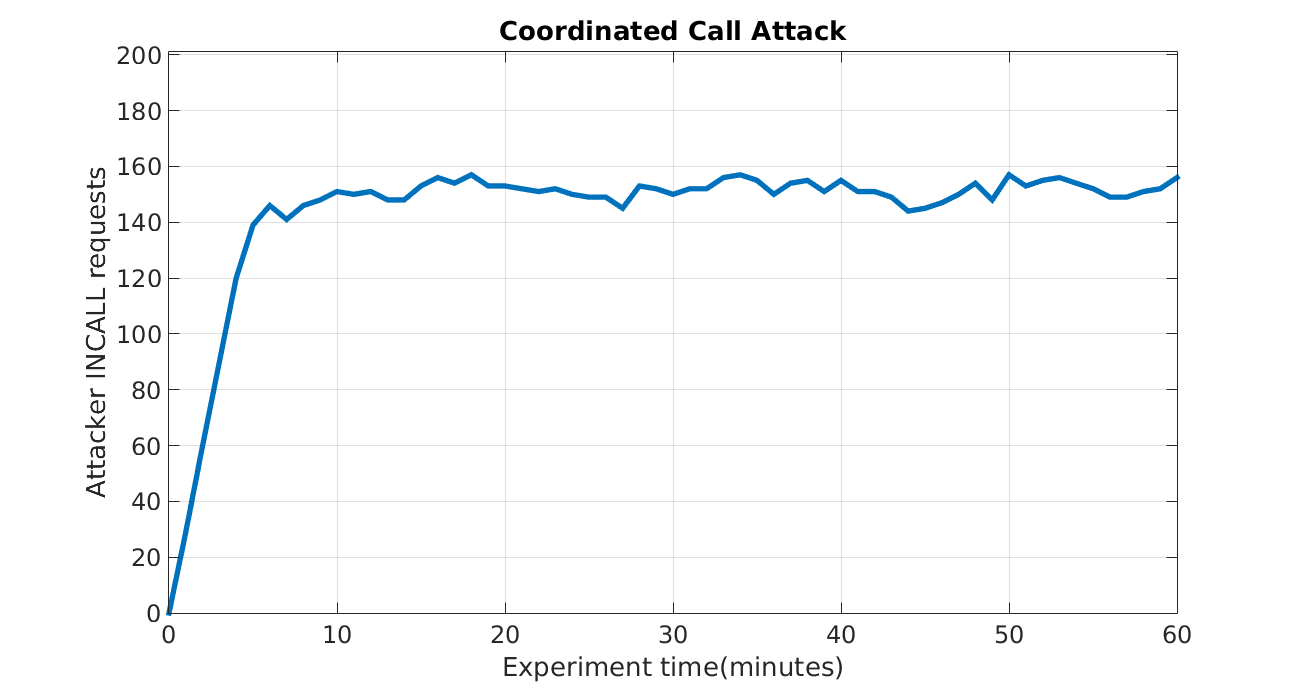}
\end{center}
\vspace{-2mm}
\caption{Attacker Call Occupancy in Buffer: Experimental Results when using \seven with a uniform dropping strategy.}
\label{fig:uni_occ}
\vspace{-2mm}
\end{figure}

We carried out experiments to test the efficiency of \seven when using a uniform dropping strategy. Our results are summarized in Figures~\ref{fig:uni-disp} and \ref{fig:uni_occ}.

The graphs depicted in Figure~\ref{fig:uni-disp} show that \seven when using a uniform dropping strategy can mitigate the \coordinatedcall attack. The results assuming a lognormal call duration is slightly better than the results when assuming an exponential call duration. In both cases, the proportion of completed calls stayed above 50\% levels even when the attacker call rate is 5 times more than the client call rate. The proportion of incomplete calls was more affected by the call duration model. Under the exponential call duration assumption, the proportion of incomplete call was around 35\% of all legitimate calls, while under the lognormal call duration assumption, the proportion of incomplete calls was of around 28\%. The proportion of unsuccessful calls stays below 10\% under both assumptions of call duration. 
We also measured the average time of incomplete calls, that is, the proportion of time that incomplete calls were able to stay in a call before they were dropped by the \seven defense strategy. When assuming both an exponential call duration and a lognormal call duration, the incomplete calls were in average around 40\% of the expected call time.

Figure~\ref{fig:uni_occ} illustrates how the attacker is able to occupy the resources of the server. While when not running \seven the attacker was able to occupy all the server's resources (Figure~\ref{fig:no_def_occ}), when using \seven with a uniform dropping strategy, the attacker is only able to occupy around 70\% of the server's resources. This may seem to be a high value, but the graph hides the fact that attackers are dropped by the defense strategy and thus the high levels of availability obtained.

\FloatBarrier

\FloatBarrier

\subsubsection{Roulette Defense}

\begin{figure}[h]
\subfigure[Exponential Call Duration.]{
  \centering
\includegraphics[width=0.55\textwidth]{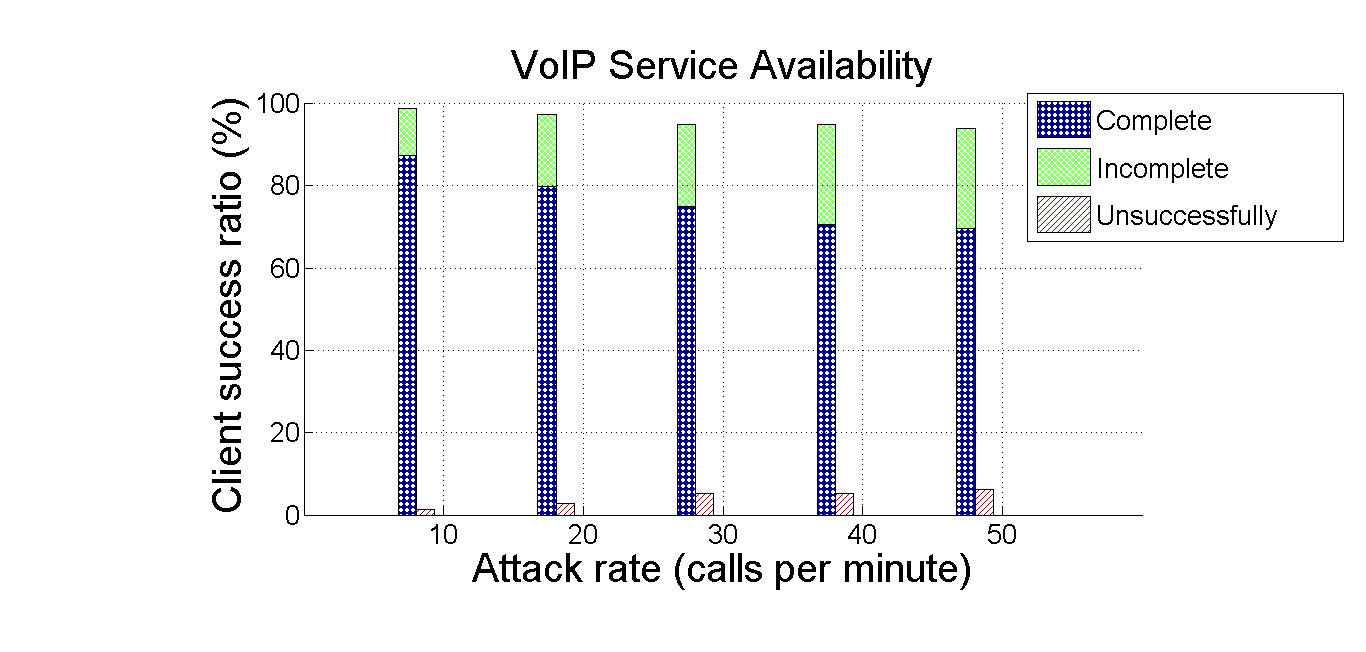}
\label{fig:exp_roul}
}
\hspace{-22mm}
\subfigure[Lognormal Call Duration.]{
  \centering
\includegraphics[width=0.55\textwidth]{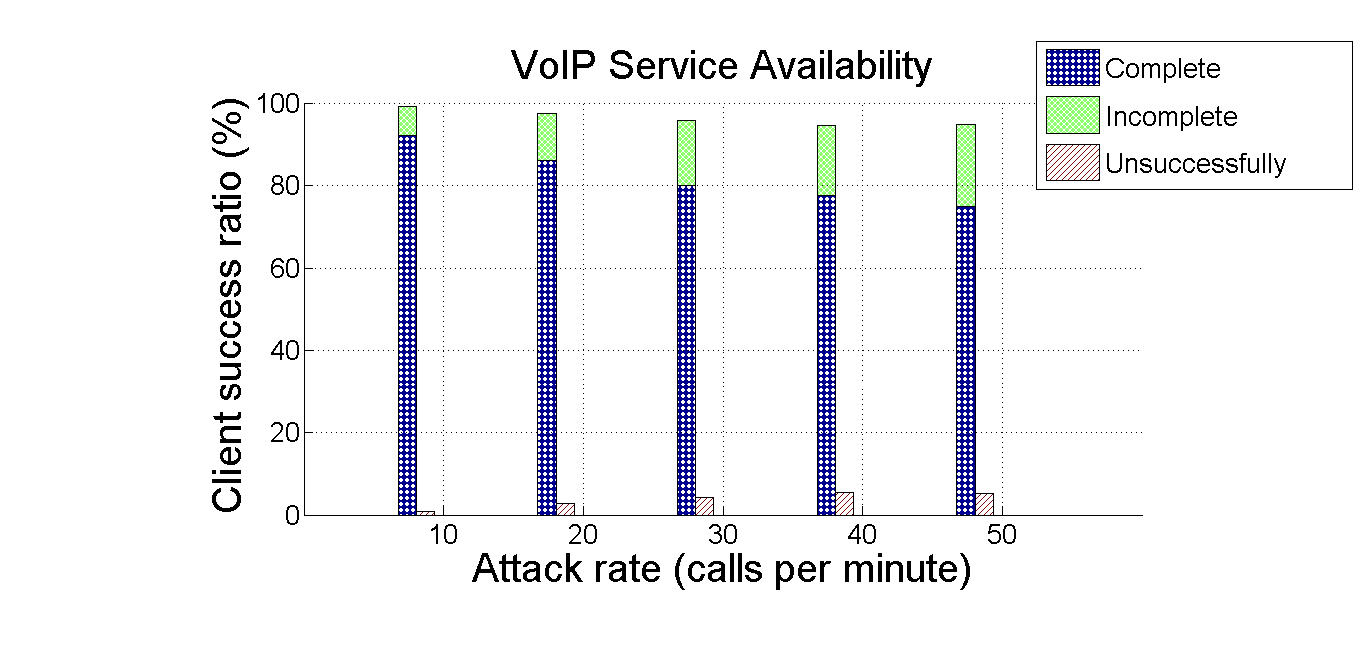}
\label{fig:log_roul}
}
\subfigure[Exponential Call Duration.]{
  \centering
\includegraphics[width=0.55\textwidth]{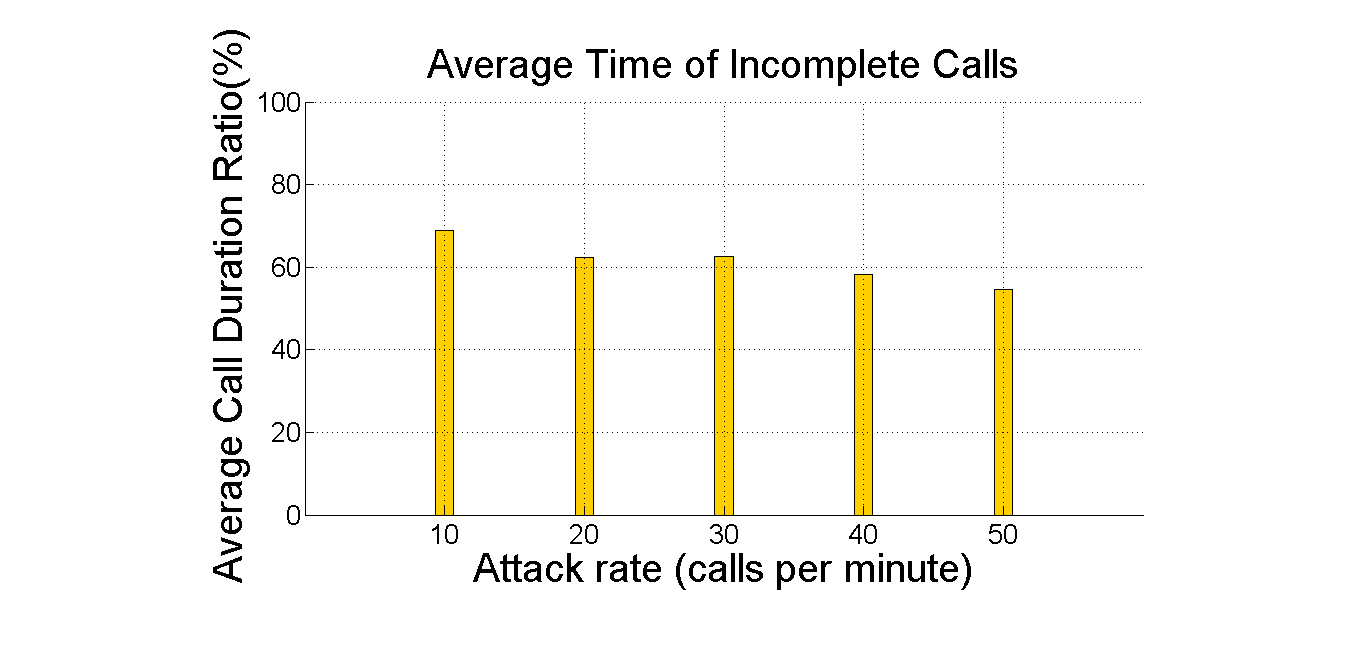}
\label{fig:exp_roul_inc}
}
\hspace{-12mm}
\subfigure[Lognormal Call Duration.]{
  \centering
\includegraphics[width=0.55\textwidth]{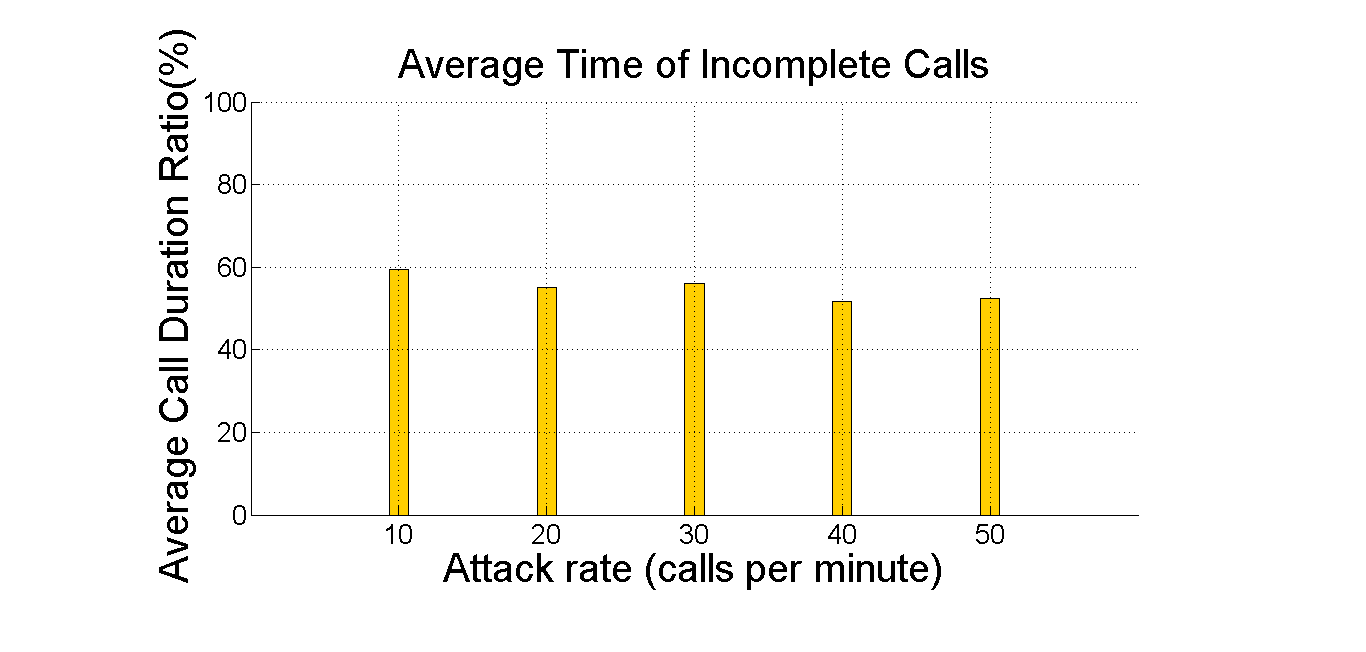}
\label{fig:log_roul_inc}
}
\caption{Client Success Ratio and Average Time of Incomplete Calls: Experimental Results when using \seven with a roulette dropping strategy.}
\label{fig:roul_disp}
\end{figure}

\begin{figure}[h]
\begin{center}
\includegraphics[width=0.66\textwidth]{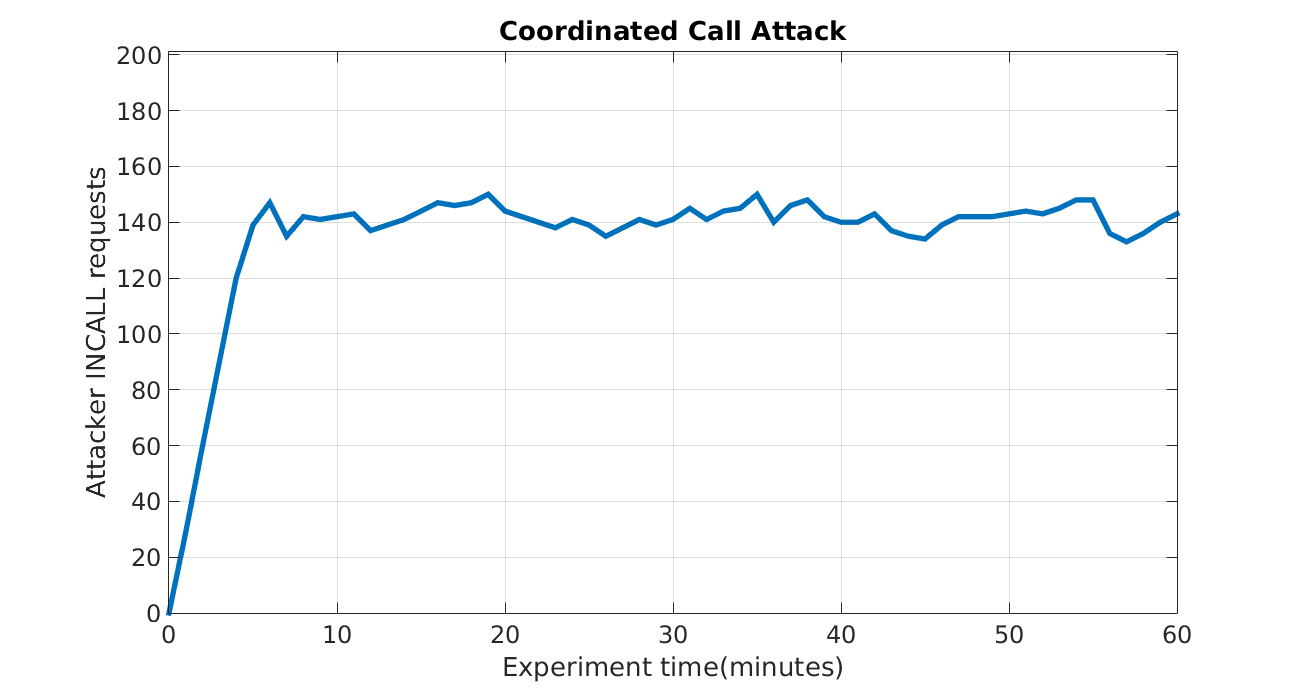}
\end{center}
\vspace{-2mm}
\caption{Attacker Call Occupancy in Buffer: Experimental Results when using \seven with a roulette dropping strategy.}
\label{fig:roul-occ}
\vspace{-2mm}
\end{figure}

Figures~\ref{fig:roul_disp} and \ref{fig:roul-occ} depict our experimental results when using \seven with the roulette dropping strategy. As with the uniform dropping strategy, the defense was able to mitigate the \coordinatedcall attack. Furthermore, the roulette strategy performed better than the uniform strategy. 

The availability depicted in Figure~\ref{fig:roul_disp} remained at high levels under both assumptions on call duration (exponential and lognormal). The proportion of completed calls was above 70\% percent when assuming an exponential call duration and above 75\% when assuming a lognormal call duration. In both cases, the proportion of incomplete calls was less than 6\%. We also measured the average time of incomplete calls. They show that these calls were able to communicate for more than 50\% of the expected time. These results are better than the results obtained using a uniform dropping strategy. 

Despite the availability results using the roulette strategy being better than the availability results obtained using the uniform strategy, the attacker was still able to occupy a similar proportion of the server's resource as depicted in Figure~\ref{fig:roul-occ}. It occupied at most 70\% of the server's resources. Intuitively, the difference in the availability between the uniform and roulette strategies is because the roulette strategy tends to drop calls with greater duration. This means that the attacker calls are more likely to be selected leaving more chance for a legitimate call to access the service.

\FloatBarrier


\subsubsection{100-Tournament Defense}

\begin{figure}[h]
\subfigure[Exponential Call Duration.]{
  \centering
\includegraphics[width=0.55\textwidth]{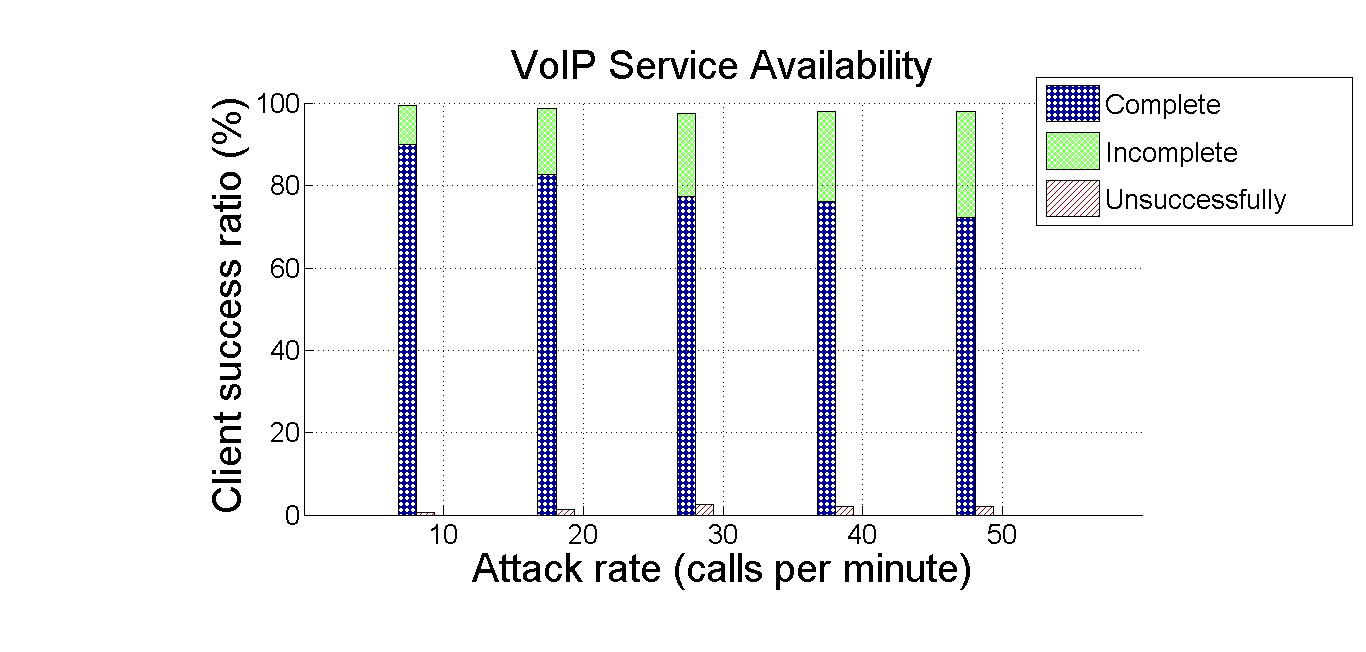}
\label{fig:exp_tour}
}
\hspace{-22mm}
\subfigure[Lognormal Call Duration.]{
  \centering
\includegraphics[width=0.55\textwidth]{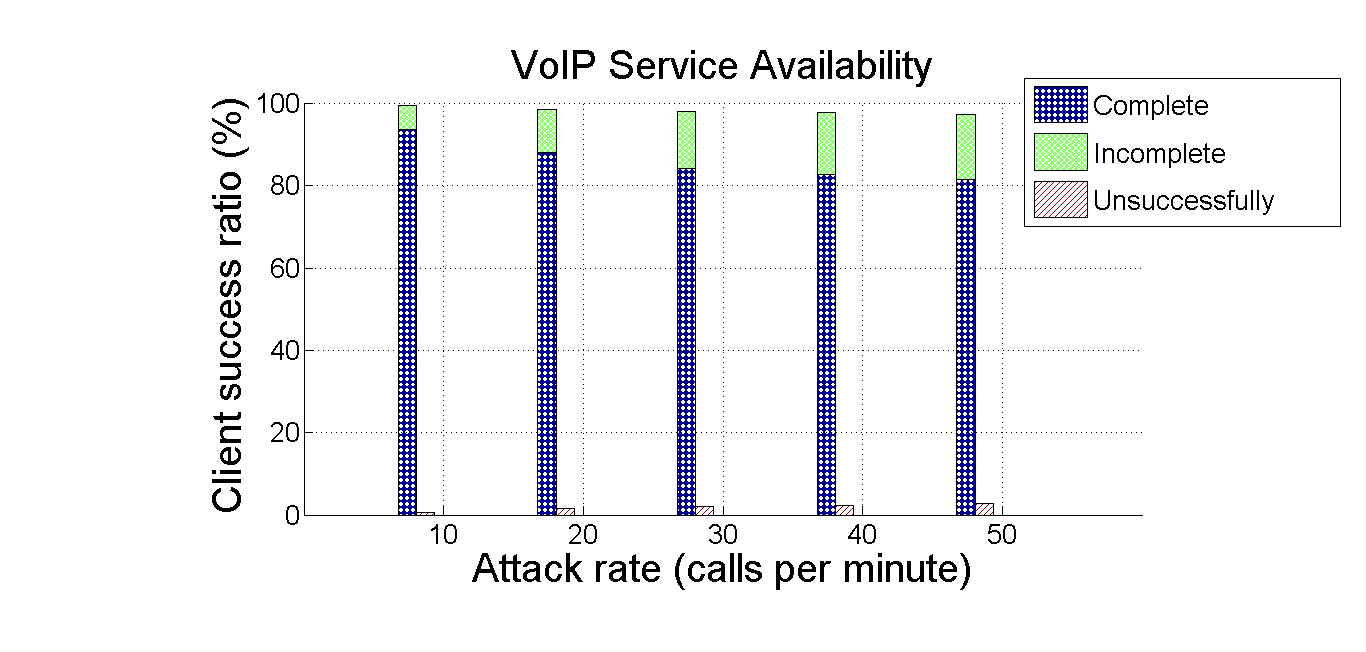}
\label{fig:log_tour}
}
\subfigure[Exponential Call Duration.]{
  \centering
\includegraphics[width=0.55\textwidth]{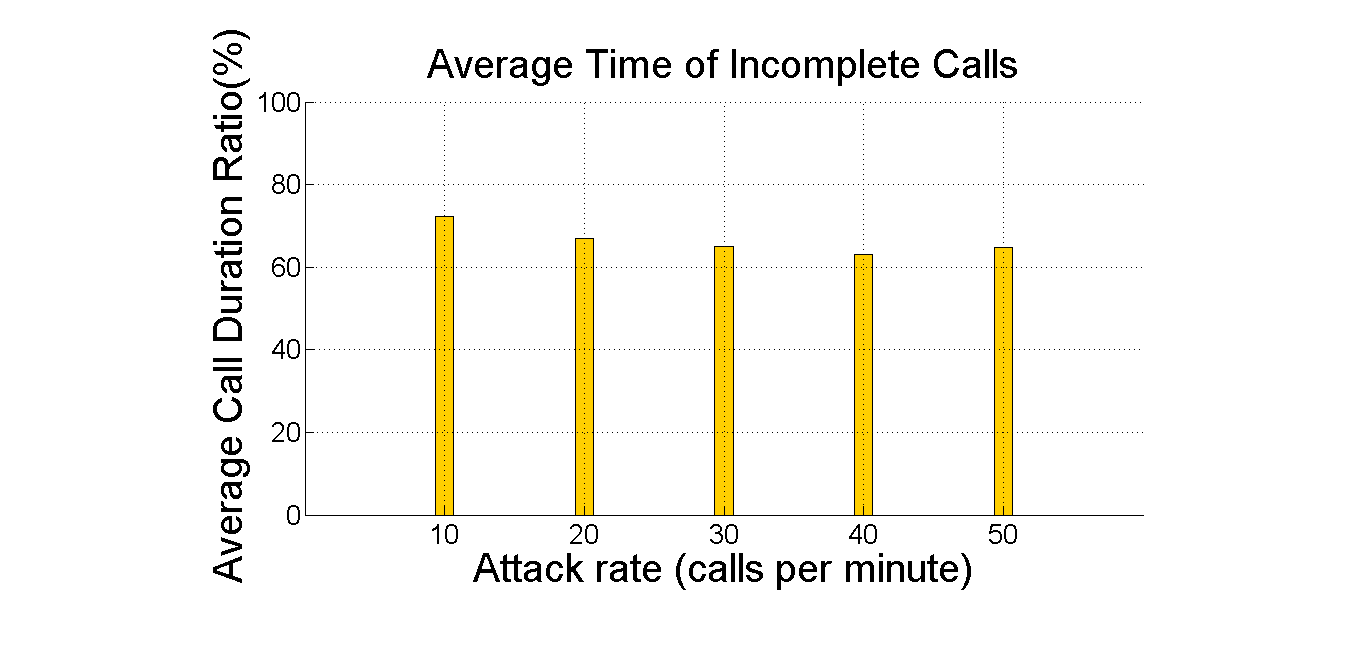}
\label{fig:exp_tour_inc}
}
\hspace{-12mm}
\subfigure[Lognormal Call Duration.]{
  \centering
\includegraphics[width=0.55\textwidth]{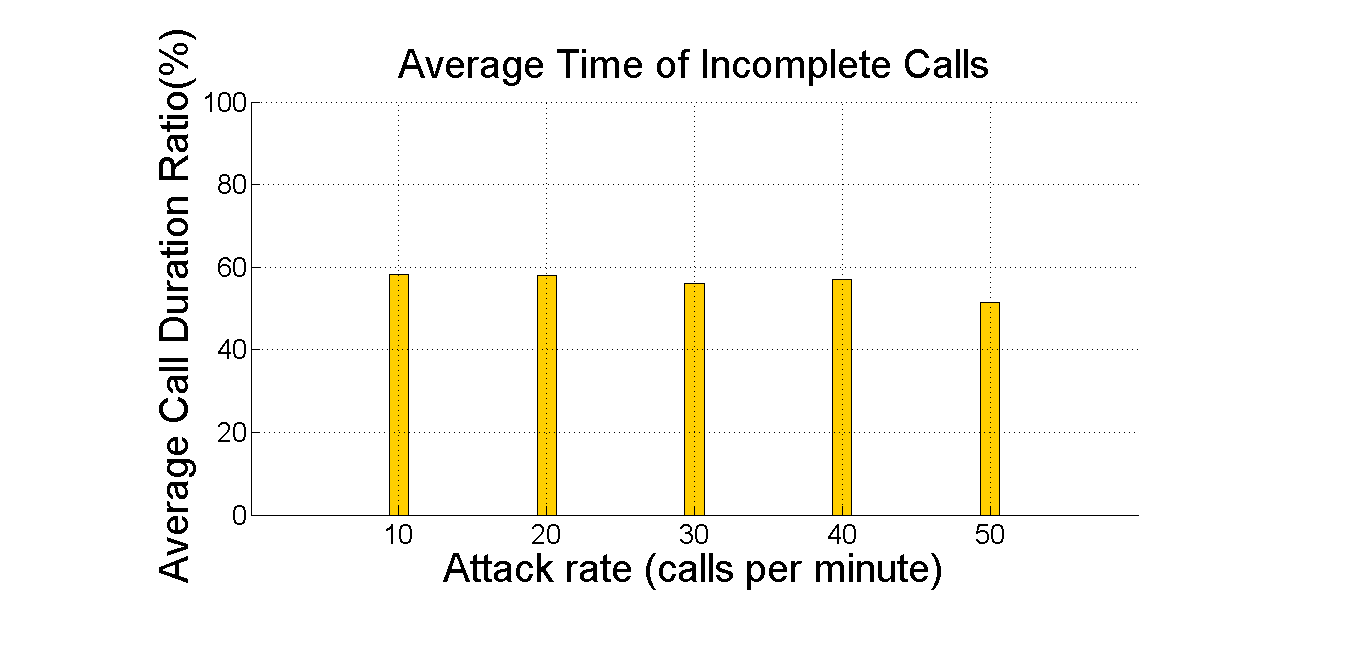}
\label{fig:log_tour_inc}
}
\caption{Client Success Ratio and Average Time of Incomplete Calls: Experimental Results when using \seven with a 100-tournament dropping strategy.}
\label{fig:tour_disp}
\end{figure}

\begin{figure}[t]
\begin{center}
\includegraphics[width=0.66\textwidth]{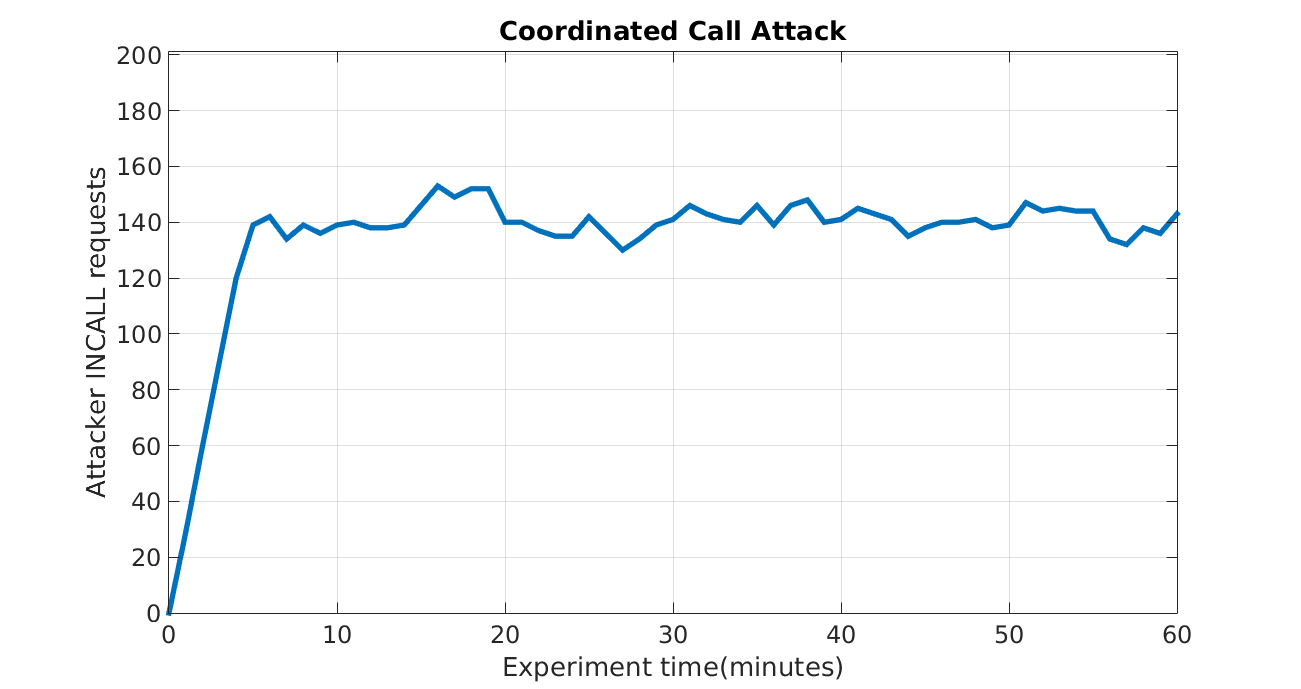}
\end{center}
\vspace{-2mm}
\caption{Attacker Call Occupancy in Buffer: Experimental Results when using \seven with a 100-tournament dropping strategy.}
\label{fig:tour-occ}
\vspace{-2mm}
\end{figure}

Our last set of experiments evaluated the efficiency of \seven with a 100-tournament dropping strategy. Figures~\ref{fig:tour_disp} and \ref{fig:tour-occ} depict our main results. The 100-tournament dropping strategy resulted in the best results when compared with the uniform and roulette strategies. 

The attacker was still able to use roughly the same amount of resources of the server as when using the uniform and roulette strategy, namely around 70\% of its resources as depicted in Figure~\ref{fig:roul-occ}. Moreover, the availability results depicted in Figure~\ref{fig:tour_disp} were slightly better than the results obtained with the roulette strategy (Figure~\ref{fig:roul_disp}). In both assumptions of call duration (following an exponential and a lognormal distributions), \seven was able to maintain high levels of availability. More than 70\% (respectively, 80\%) of calls were completed when assuming call duration following an exponential distribution (respectively, lognormal distribution). The proportion of incomplete calls reached levels around 25\% of all calls when assuming an exponential cal duration and reached 16\% of all calls when assuming lognormal call duration. Thus, more than 95\% of all legitimate calls were able to reach the incall status, which means that they were able to communicate.

Finally, incomplete calls were interrupted by \seven after communicating more than 60\% of the expected time when assuming exponential call duration and more than 50\% when assuming lognormal call duration.

\FloatBarrier

\subsection{Impact of Using \seven\ When not Suffering an Attack}

We investigate additionally the impact of using \seven\ as a proxy and implementing 
the selective strategy on the performance of Asterisk. Asterisk modules provide many statistics on the performance of the system. The following tables contain the number of requests according to the time intervals for the time to respond (TTR) when using and not using \seven\ during normal situation, that is, when only receiving legitimate calls.         

\begin{small}
\begin{center}
\begin{tabular}{@{~}c@{~}c@{~}c@{~}c@{~}c@{~}c}
\toprule
\multicolumn{6}{c}{\textbf{Not Using \seven}}\\
\toprule
\textbf{TTR (ms)} & $[0,1]$ & [3,4] & [4,5] & [8,9] & [10,20]  \\
\midrule
\textbf{Num Requests} & 1837 & 1 & 960 & 2 & 2 \\
\bottomrule
\end{tabular}
\end{center}  
\end{small}

\begin{small}
\begin{center}
\begin{tabular}{@{~}c@{~}c@{~}c@{~}c@{~}c@{~}c@{~}c@{~}c@{~}c@{~}c@{~}c@{~}}
\toprule
\multicolumn{11}{c}{\textbf{Using \seven}}\\
\toprule
\textbf{TTR (ms)} & $[0,1]$ & [3,4] & [4,5] & [7,8] & [10,20] & [20,30] & [30,40] & [40,50] & [50,100] & [100,150]  \\
\midrule
\textbf{Num Requests} & 19 & 2 & 404 & 341 & 657 & 434 & 148 & 96  & 70 & 8 \\
\bottomrule
\end{tabular}
\end{center}  
\end{small}

As one can observe, there is an impact to TTR when using \seven. While without \seven\ most of the requests are responded within 5ms, with \seven, most of the requests are responded within 100ms. 

Such a delay does not greatly impact user experience as 100 ms is negligible with respect to the time users wait until establishing a call, \eg, waiting until Bob accepts the call which normally takes some seconds to happen. Moreover, as \seven\ only acts on SIP messages (Initiation and Termination phases of Figure~\ref{fig:sip-normal}), \seven does not affect user experience when the parties are in a call (Communication Phase in Figure~\ref{fig:sip-normal}). 

Finally, it seems possible to improve \seven's performance by improving our implementation, \eg, implementing it as a module instead of a proxy. This is left, however, to future work.

%% file: structure/comparison.tex

\subsection{Differences between Simulations and Experiments}

There are some important differences between the formal model and the experimental set-up. For a starter, the formal specification abstracts several aspects present in the experimental set-up. For example:
\begin{itemize}
  \item We did not model how Asterisk actually manages its workers/threads. Asterisk has a number of modules that among other things, maintain call statistics, convert calls encoded some codex to another, etc;
  \item In our experiments, there are other applications running in parallel with Asterisk that have to compete for resources (CPU and network interface for example). Our formal model does not incorporate this;
  \item We use a simplified model for network latency with constant latency time;
  \item While the parameters, \eg, $k$, incoming client and attacker traffic, used in the simulations were proportional to the parameters used in the experiments, they were much lower to the ones used in the experimental results. If we used the actual values for these parameters, simulations would have taken much longer;
  \item In our experiments, \seven\ is used as a proxy, while in our formal model the defense was incorporated into the application.   
\end{itemize}

Despite these important differences/abstractions, as we compare in more detail next, the simulation results corresponded to many of the experimental results. For example in terms of availability, \ie, proportion of completed, incomplete and unsuccesful calls, the simulation results correctly indicated the power of the attack and the efficiency of \seven\ mitigating this attack. They also correctly indicated which dropping strategy is better and how availability changes with the increase on the attack rate. The simulation results were less accurate in predicting the time of incomplete calls reaching a difference of 30\%. The reasons for this discrepancy are not completely clear, but we believe it has to do with the abstractions mentioned above. We leave this investigation to future work.

\subsection{Detailed Comparisons}

Typically each simulation takes about 30 seconds to be completed. In contrast, each experiment carried out on the network took 60 minutes. This means that specifiers can quickly test different selective strategies using formal verification before implementing the necessary machinery and carrying out experiments. Once a selective strategy is shown by formal verification to have reasonable results, experiments can be carried out to validate the chosen defense. 

In this section, we compare the results obtained through formal verification detailed in Section~\ref{sec:sim} and the results obtained by carrying out experiments detailed in Section~\ref{sec:exp}. In general, availability results obtained through formal verification indeed corresponded to our experimental results showing a high degree of accuracy for our simulation results. 

\begin{figure}[h]
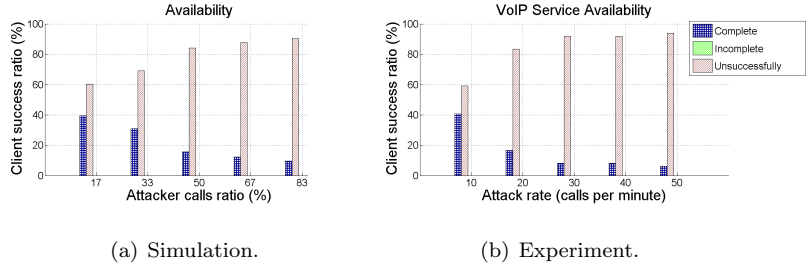

\subfigure[Simulation.]{
  \centering
\includegraphics[width=0.55\textwidth]{figures-simulations/no_defense_log.png}
}
\hspace{-21.5mm}
\subfigure[Experiment.]{
  \centering
\includegraphics[width=0.55\textwidth]{figures-coordinated/no_defense_log.png}
}
\caption{Client Success Ratio: Comparison between simulation and experimental results for lognormal call duration.}
\label{fig:comparison_no_def}
\end{figure}

\paragraph{Efficiency of the \coordinatedcall Attack:}
Our simulation results with the scenario without defense showed that the \coordinatedcall is effective if the server does not have any defense mechanism under both assumptions of duration calls (exponential and lognormal). This result was also observed in our experimental results. Figure~\ref{fig:comparison_no_def} on the service availability illustrates this correspondence.

\begin{figure}[h]
\subfigure[Simulation.]{
  \centering
\includegraphics[width=0.55\textwidth]{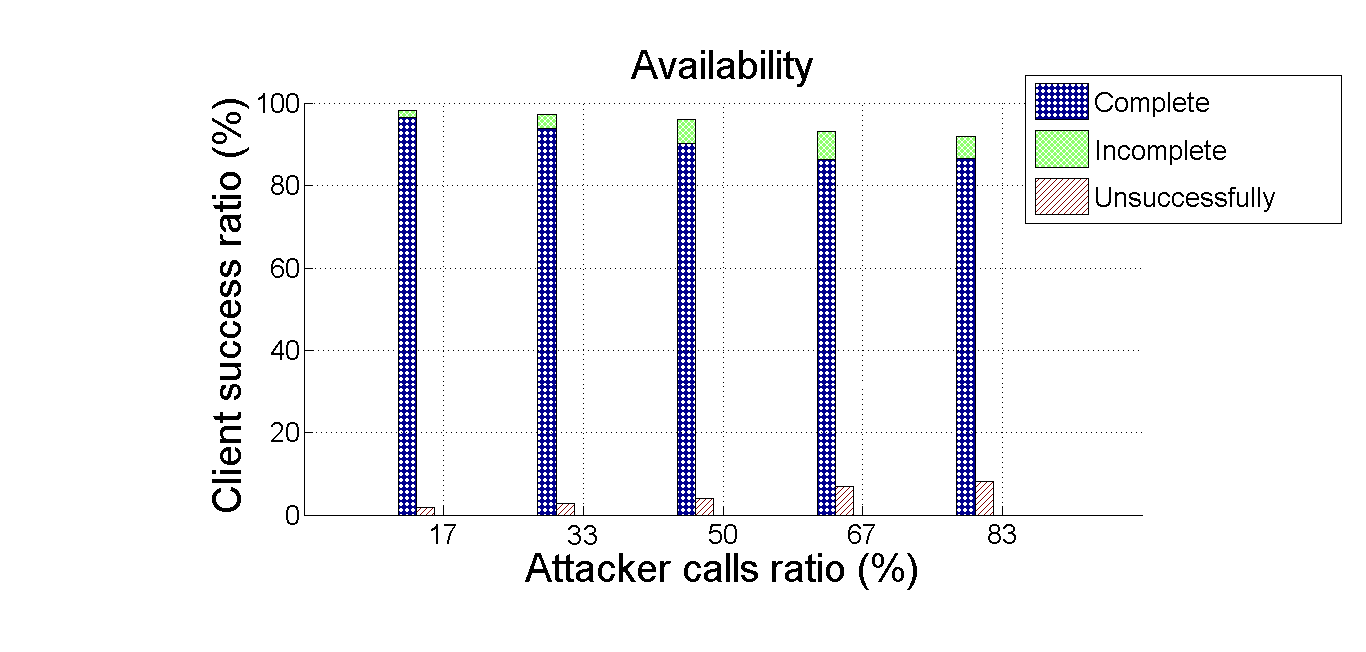}
}
\hspace{-21.5mm}
\subfigure[Experiment.]{
  \centering
\includegraphics[width=0.55\textwidth]{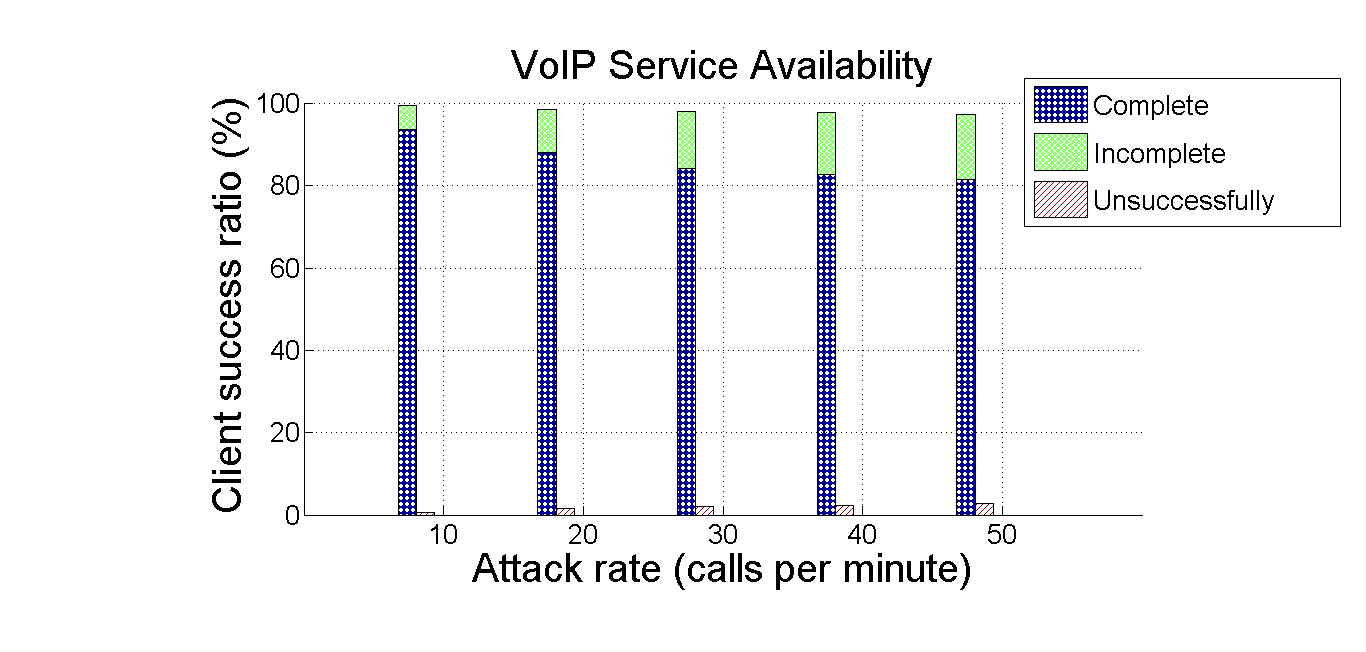}
}
\caption{Client Success Ratio: Comparison between simulation and experimental results for lognormal call duration when using \seven with a tournament dropping strategy.}
\label{fig:comp_lognormal}
\end{figure}

\paragraph{Efficiency of \seven:} 
All our simulations results using scenarios with \seven indicated that \seven is indeed a good defense for mitigating the \coordinatedcall attack. The greater proportion of calls were completed calls, while a smaller proportion of the calls were incomplete and a minority of calls were unsuccessful. The same behavior was observed by our experiments. This is illustrated by Figure~\ref{fig:comp_lognormal}.

\paragraph{Dropping Strategies Evaluation:} Our simulations were able to predict that the tournament dropping strategy would perform best. However, it was not able to forecast that the roulette strategy would perform better than the uniform strategy. It is not clear to us why this was the case. 

\paragraph{Better Performance for Lognormal Call Duration:} 
Our simulations results also predicted that selective strategies would perform better in scenarios where call duration of legitimate clients follows a lognormal distribution, such as in VoIP communication, than scenarios with exponential call duration, such as in traditional telephony. 
This was the case independent on the dropping strategy used (uniform, roulette or $\frac{k}{2}$-tournament). The same behavior was observed in our experimental results. 

\begin{figure}[h]
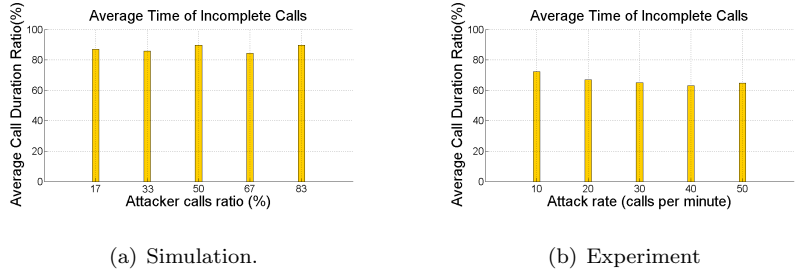

\subfigure[Simulation.]{
  \centering
\includegraphics[width=0.55\textwidth]{figures-simulations/tournament_log_incomplete.png}
}
\hspace{-12mm}
\subfigure[Experiment]{
  \centering
\includegraphics[width=0.55\textwidth]{figures-coordinated/tournament_exp_incomplete.png}
}
\caption{Average Time of Incomplete Calls: Comparison between simulation and experimental results when using \seven with the $\frac{k}{2}$-tournament dropping strategy.}
\label{fig:compa_avg_time}
\end{figure}

\paragraph{Average time of Incomplete Calls:} Our simulations results also indicated that the average time that incomplete calls stayed communicating before being drop by \seven was relatively high: more than 80\% of the expected time when using the tournament strategy under a lognormal call duration. This results diverged from the experimental results which observed an average time of incomplete calls of around 60\%. This is illustrated by Figure~\ref{fig:compa_avg_time}. 

\FloatBarrier

However, for other scenarios, the difference between simulation and experimental results was greater reaching 30\% of difference. It is not clear to us what are the reasons for this difference. We suspect that the modeling of the time delays should be improved. In any case, our results suggest that while simulations provide quite accurate results on availability, it is less accurate on specific timing analysis. We observed a similar behavior during our experiments and simulations when modeling our defense for mitigating Application-Layer DDoS attacks~\cite{dantas14eisic}.

%% file: structure/future-work.tex
This paper formalized a new selective defense, called \seven, for mitigating \coordinatedcall\  attacks. We have shown that using state-dependent probability distributions for selecting which calls are to be processed results in high levels of availability. We proposed three defenses based on the dropping strategy method (uniform, roulette and $\frac{k}{2}$-tournament). We carried out simulations and experiments assuming traditional telephony and VoIP communications. In both cases, we observed that our \seven was able to mitigate the \coordinatedcall attack. Finally, we compared the results obtained using our formal analysis with the results obtained by experimentation obtaining a high accuracy. This further supports the value of formal analysis during the development of selective defenses for mitigating DoS attacks.

Most of the existing work~\cite{Ha:2009:DIS:1655925.1656137,5426267,6195475,6720187,5541813,5630386} on mitigating DoS attacks on VoIP services focuses on flooding attacks, such as the SIP-Flooding attack. They analyze the network traffic and whenever they observe an abrupt increase in the traffic load, they activate their defenses. The network traffic is usually modeled using some statistical approach, such as correlating the number of INVITE requests and the number of requests that completed the SIP initiation phase~\cite{Ha:2009:DIS:1655925.1656137} or using more complicated metrics such as Helling distance to monitor traffic probability distributions~\cite{5426267,6195475,6720187}. Other solutions place a lower priority on INVITE messages, which are only processed when there are no other types of request to be processed~\cite{5541813,5630386}.

As the Coordinated Call Attack emulates legitimate client traffic not causing an unexpected sudden increase in traffic, all these defenses are not effective in mitigating the Coordinated Call Attack. The few solutions we found in the literature for this type of attack are commercial tools that act as a firewall which monitor all the call traffic and the signaling~\cite{securelogix,TransNexus} or analyze audio samples~\cite{pindrop} in order to differentiate the fraudulent calls from the legitimate ones. Less sophisticated mechanisms~\cite{CUBE} monitors all the incoming requests and reject those whose IPs do not belong to a list of trusted IPs. Clearly such approaches does not work well when the attackers are malicious users whose IPs are in the trusted list and are not using automation to make the calls. In addition, these commercial tools can be expensive for small businesses to purchase and maintain, and they require technical expertise for proper installation.

One main advantage of our proposed solution is that it is not tailored using many specific assumptions on type of service. The only assumption used is a previous knowledge of the average call duration, which can be easily inferred from the service call history. Moreover, our solution can be easily integrated with other mechanisms such as the IP filtering approach used in~\cite{CUBE}.  

\cite{huici09globecom} proposes a filtering mechanism for SIP flooding attacks. It is not clear whether such mechanisms will be enough for mitigating the Coordinated VoIP attack, as the number of messages needed to carry out such attack is much less. Wu \etal~\cite{wu04dsn} have proposed a mechanism to identify intruders using SIP by analyzing the traffic data. Although we do not tackle the identification of intruders problem, we find it an interesting future direction.

The formalization of DDoS attacks and their defenses has been subject of other papers. For example, Meadows proposed a cost-based model in \cite{meadows99csfw}, while others  use branching temporal logics~\cite{mahimkar05csfw}. This paper takes the approach used in \cite{alTurki09entcs,eckhardt12wadt,eckhardt12fase}, where one formalizes the system in Maude and uses the Statistical Model Checker PVeStA to carry out analyses. While~\cite{alTurki09entcs,eckhardt12wadt,eckhardt12fase} modeled traditional DDoS attacks exploiting stateless protocols on the transport/network layers, we are modeling stateful Application Layer DDoS attacks. Moreover, the quality measures used for VoIP services under TDoS attacks, described in Section~\ref{sec:seven}, are different to the quality measures considered in the previous work. 

More recently~\cite{dantas14eisic}, we proposed \seven\ showing that it can be used to mitigate \ADDoS\ attacks that exploit the HTTP protocol. This paper shows that \seven\ can also be used to mitigate DDoS attacks in VoIP protocols, but in order to do so one needs state-dependent probabilistic distributions. This is because of the quality requirements that we need in VoIP communications. We would like to give a priority to the types of call that should be given more chances to keep using resources of the server. In particular, we give preference to calls that do not take more than the average duration time. Such quality measures are not present in HTTP protocols that we analyzed in \cite{dantas14eisic}. 

For future work, we are going to investigate the mitigating of other types of attacks, such as volumetric and amplification attacks. We are also thinking on intrusion detection mechanisms. We are also interested in building defenses for mitigating amplification attacks~\cite{shankesi09esorics}. We have also been using \seven\ for mitigating High-Rate ADDoS attacks using Software Defined Networks~\cite{henrique16sbrt}. We are also investigating ways to improve simulation accuracy by improving the modeling of timing aspects of the system, such as processing and network delay.